\documentclass[prx,twocolumn,aps,superscriptaddress,showpacs,amsmath,amssymb,footnoteinbib]{revtex4-1}

\usepackage[utf8]{inputenc}
\usepackage{graphicx}
\usepackage{latexsym}
\usepackage{amsmath}
\usepackage{amssymb}
\usepackage{amsfonts}
\usepackage{color}

\usepackage{amssymb}
\usepackage{bm}
\usepackage{verbatim}

\usepackage[colorlinks,bookmarks=false,citecolor=blue,linkcolor=red,urlcolor=blue]{hyperref}
\usepackage[all]{hypcap} 

\usepackage{color}
\usepackage[all]{hypcap} 

\usepackage{soul}
\usepackage{ulem} 

\begin{document}

\title{Electron-phonon heat transfer in giant vortex states}

\author{A.~V.~Samokhvalov}
\affiliation{Institute for Physics of Microstructures, Russian Academy of Sciences, Nizhny Novgorod, Russia}
\affiliation{Lobachevsky State University of Nizhni Novgorod, 23 Prospekt Gagarina, 603950 Nizhni Novgorod, Russia}
\author{I.~A.~Shereshevskii}
\affiliation{Institute for Physics of Microstructures, Russian Academy of Sciences, Nizhny Novgorod, Russia}
\author{N.~K.~Vdovicheva}
\affiliation{Institute for Physics of Microstructures, Russian Academy of Sciences, Nizhny Novgorod, Russia}
\author{M.~Taupin}
\affiliation{Institute of Solid State Physics, Vienna University of Technology, Wiedner Hauptstrasse 8-10, 1040 Vienna, Austria}
\author{I.~M.~Khaymovich}
\affiliation{Institute for Physics of Microstructures, Russian Academy of Sciences, Nizhny Novgorod, Russia}
\affiliation{Max Planck Institute for the Physics of Complex Systems, N\"{o}thnitzer Str. 38, 01187 Dresden}
\author{A.~S.~Mel'nikov}
\affiliation{Institute for Physics of Microstructures, Russian Academy of Sciences, Nizhny Novgorod, Russia}
\affiliation{Lobachevsky State University of Nizhni Novgorod, 23 Prospekt Gagarina, 603950 Nizhni Novgorod, Russia}


\begin{abstract}
We examine energy relaxation of non-equilibrium quasiparticles in
different vortex configurations in ``dirty'' $s$-wave
superconductors. The heat flow from the electronic subsystem to phonons in a mesoscopic
superconducting disk with a radius of the order of several
coherence lengths is calculated both in the Meissner and giant
vortex states using the Usadel approach.
The recombination process
is shown to be strongly affected by interplay of the subgap
states, located in the vortex core and in the region at the sample
edge where the spectral gap $E_{\rm g}$ is reduced by the Meissner
currents.
In order to uncover physical origin of the results, we develop a semiquantitative analytical approximation
based on the combination of homogeneous solutions of Usadel equations in Meissner and vortex states of a mesoscopic superconducting disc
and analytically calculate the corresponding spatially resolved electron-phonon heat rates.
Our approach provides an important information about
non-equilibrium quasiparticles cooling by the magnetic-field
induced traps in various mesoscopic superconducting devices.
\end{abstract}
%

\maketitle

\section{Introduction}\label{IntroSection}
The progress in cryogenic superconducting (SC) devices and
technologies, such as electronic refrigerators and thermometers~\cite{Giazotto-Pekola-rmp06}, radiation detectors~\cite{Bolometers1,Bolometers2}, and qubit systems for quantum
information processing~\cite{Qubit1,Qubit2} requires an improved
understanding of the quasiparticle thermalization mechanisms. Such
mesoscopic devices with nanoscale dimensions operating at
subkelvin temperatures are easily driven out of equilibrium via
processes involving electromagnetic field absorption, hot
quasiparticle injection, or the operational drive. As a result,
there is a significant concentration of non-equilibrium
quasiparticles (QPs), present in a driven superconductor even for
the temperatures far away from the drive location well below the
temperature of superconducting transition~\cite{Catelani-Pekola_review21}. Typically, the presence
of excess QPs  is characterized by an effective electron
temperature $T$, which exceeds the temperature of the phonon bath
$T_{\rm ph}$. Excess QPs destroy coherence of qubit systems~\cite{QubitDecog1,QubitDecog2,QubitDecog3}, decrease the quality
factor of superconducting resonators~\cite{Q-Resonat1,Q-Resonat2},
reduce the efficiency of cooling in electronic refrigerators~\cite{Refrig1,Refrig2}, and result in the excess current in single
electron~\cite{TurnStile,SQS_turnstile,NEq_turnstile} and heat quantum~\cite{heat_turnstile} turnstiles.
The energy-relaxation rate of the non-equilibrium QPs is also known to affect the
characteristics of detectors of electromagnetic radiation~\cite{deVisser_detectors,Catelani-Pekola_review21}. Thus,
the overheating of QPs or unwanted population in general appear to
be major factors limiting the performance of low temperature SC
devices. This is why the study of the mechanisms of
non-equilibrium QPs relaxation in superconductors of mesoscopic
dimensions seems to be rather important (see e.g.,~\cite{Arutyunov-jpcm18} for review).

In order to suppress quasiparticle poisoning and to prevent
overheating, density of non-equilibrium QPs in a device should be
reduced. Simple lowering the phonon-bath temperature $T_{\rm ph}$
often is not sufficient, and to remove QPs in superconducting
devices additional quasiparticle traps should be used: either
normal-metal films connected to the superconductor~\cite{QP_traps1a,QP_traps1b,QP_traps1c}, Andreev bound states in
weak links~\cite{QP_traps_An}, or regions with reduced or
destroyed SC gap~\cite{QP_traps2a,QP_traps2b,QP_traps2c,QP_traps2d}. In this case
QPs are trapped by the region with no energy gap (or suppressed
gap) away from the active region. One of the possible types of QPs
traps can be formed by regions with the suppressed SC gap, that
appear in the Meissner and vortex states and can be successfully
controlled by the external magnetic field~\cite{QP_traps3a,QP_traps3b,QP_traps3c,QP_traps3d,QP_traps3e,QP_traps4a,Nakamura17,Taupin-Khaymovich-NatCom16,Catelani-Pekola_review21,Ueda21}.
This method has a number of advantages due to using of the same
material as the rest of the device, since perfect matching of
different parts of the device without barriers or interface
potentials is provided. Besides, magnetic field allows to tune the
trap controllably. The controllable use of such traps in various
applications mentioned above assumes, certainly, understanding of
their cooling capacities.

The hot electrons dissipate the heat typically via the interaction
with the phonons~\cite{Eliashberg-JETP72,Kaplan-Scalapino-PRB76,Kopnin-NeqSc}. In
three dimensional normal metals of volume $\mathcal{V}$ (when the mean free path $\ell$ is small compared to all other length scales) the heat
flow $\dot{Q}_N$ related to the electron--phonon (e-ph) relaxation
is determined by the well--known relation
\begin{equation}\label{Elec-Phon-Normal}
    \dot{Q}_N = \Sigma \mathcal{V} \left( T^5 - T_{\rm ph}^5 \right)\,,
\end{equation}
where $\Sigma$ is the material constant for electron--phonon
coupling~\cite{Giazotto-Pekola-rmp06}. Quasiparticle recombination
in a bulk superconductor with an $s-$wave gap via coupling to the
phonons has been widely studied~\cite{Wellstood-prb94,Timofeev-prl09,Maisi-prl13,Kopnin-NeqSc,Bergeret-Silaev-rmp18}.
The hard energy gap $E_{\rm g}$ in the quasiparticle spectrum of
the superconductor is known to suppress the recombination of the
hot QPs making the relaxation rate $\nu_{e-ph}$ to be
exponentially slow ($\sim \exp(-E_{\rm g} / k_B T)$) because QPs
need to possess an energy exceeding the gap $E_{\rm g}$ in order
to recombine. The electron--phonon relaxation remains extremely
slow even in the presence of Abrikosov vortices and subgap
excitations localized in the vortex core in clean superconductors
at ultra low temperatures~\cite{Kopnin-prb99}. The resulting
relaxation rate enters the rate equations for the concentration of
QPs and phonons which allow to get the full description of the
non-equilibrium processes in superconducting devices~\cite{Rothwarf-prl67}. The electron--phonon relaxation in
mesoscopic SC samples and SC point contacts looks rather different
compared to dynamics of bulk quasiparticles~\cite{Ivanov-Feigelman-JETPL98,Zgirski-Urbina-PRL11,Olivares-Urbina-PRB14,Savich-PRB17}.
The simplified model of quasiparticle dynamics and electron
cooling both in the Meissner and vortex states of a mesoscopic
superconductor was proposed recently in
Ref.~\onlinecite{Taupin-Khaymovich-NatCom16} to explain
experimental measurements of the characteristics of
non-equilibrium quasiparticle distributions in a mesoscopic SC
island in a single-electron transistor setup with normal metal
leads, however, that model worked well only for the Meissner and single-quantum vortex states.

The main goal of our work is to overcome this limitation by developing a quantitative description
of the electron-phonon heat transfer processes in a diffusive
mesoscopic SC disk of the size comparable to several
superconducting coherence lengths $\xi_0$ placed in external
magnetic field oriented perpendicular to the plane of the disk. In
such nanoscale samples theory predicts the existence of rather
exotic vortex configurations -- so-called "giant"\ (or
multiquantum) vortices in the disk center~\cite{Peeters-Nature97,geim}. These exotic vortex states are
formed due to the confinement effect of the screening
supercurrents and have been observed by a variety of experimental
methods~\cite{Peeters-Nature97,geim,Moshchalkov-PRB04,Moshchalkov-PRB08,Grigorieva,Kokubo-PRB10},
including scanning tunneling microscopy/spectroscopy studies~\cite{Peeters-PRL04,Roditchev-PRL09,Roditchev-PRL11,Nishio-PRL08,Moshchalkov-PRB16,Hess-PRL89}.
Electronic structure of mesoscopic superconductors is known to be
sensitive to the applied magnetic field, related vorticity and
vortex configurations~\cite{tanaka1,virt,tanaka3,Melnikov-PRL05,mrs1,mrs2,Peeters-PRL11}. 
The overall spectral characteristics and local density of states
(LDOS) of the mesoscopic disk was shown to be strongly affected by
an interplay of the subgap states, located in the vortex core and
in the region at the sample edge where the spectral gap $E_{\rm
g}$ is reduced by Meissner currents~\cite{Samokhvalov-PRB19-DOS}.
Keeping in mind the high sensitivity of e-ph relaxation to the
local density of states~\cite{Taupin-Khaymovich-NatCom16}, one can
expect a strong dependence of the total QPs heat flow and the
relaxation rate $\nu_{e-ph}$ in the mesoscopic disk on the applied
magnetic field and related vorticity.  Note that both the giant
vortex cores and the region near the sample edge with the flowing
Meissner screening currents can be clearly viewed as quasiparticle
traps created by the applied magnetic field. It is the joint
effect of contributions from the core and edge traps which determines
the total QPs heat flow in a mesoscopic sample.

The paper is organized as follows. In Sec.~\ref{ModelSection} we
briefly discuss the basic equations. In
Sec.~\ref{GiantVortexStates} we calculate the superconducting
critical temperature $T_{\rm c}$ and study the switching between
the states with different vorticity $L$ while sweeping the
magnetic field. In Sec.~\ref{E-PhHeatFlux} we calculate
numerically the electron--phonon heat flow in a mesoscopic SC disk
with a giant vortex.  In Sec.~\ref{HeatFluxHomoSect} we propose a
simplified analytical model to describe the electron--phonon heat
flow and check its applicability. We summarize our results in
Sec.~\ref{SumUpSection}.

\section{Model and basic equations}\label{ModelSection}
%

\begin{figure}[t!]
\includegraphics[width=0.35\textwidth]{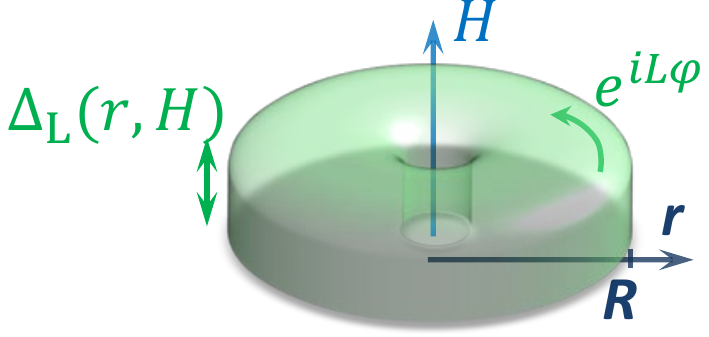}
\caption{(Color online) \textbf{Schematic picture of the spatial
order parameter distribution $\Delta_L(H)$} (shown by semi-transparent green
color) in the superconducting disk of the radius $R$ with the
giant $L$-quantum vortex, Eq.~\eqref{Delta_eq}, in the applied
perpendicular magnetic field~$H$.
}
\label{Fig0}
\end{figure}
To set the stage, here we describe the considered system and main relevant equations and approximations.
The section contains the standard notations and literature results, therefore a reader,
acquainted with this formalism and interested in the original result, can directly jump to Sec.~\ref{GiantVortexStates}.

The system in the focus is a thin mesoscopic superconducting (SC)
disk of a radius $R$ comparable to several superconducting
coherence lengths $\xi$ at a finite temperature $T$. We
concentrate on more experimentally relevant $s$-wave
superconductors in the dirty limit where the mean free path $\ell$ is
the smallest length scale in the system. In terms of the elastic
electron scattering rate $\tau^{-1}$ and the bare superconductor
transition temperature $T_{\rm cs}$ the above dirty limit
condition reads as follows $k_B T_{\rm cs} \tau/\hbar \ll 1$. In order to
generate vortices that trap quasiparticles in their cores, an
external magnetic field $\mathbf{H} = H \mathbf{z}_0$,
perpendicular to the SC disk plane, is applied (Fig.~\ref{Fig0}).
We focus on the experimentally relevant situation of a small disk
thickness $d$ and disregard the magnetic field, induced by
Meissner supercurrents in the disk. This approximation holds as
soon as $d$ is small compared to the London penetration depth
$\lambda$ in such a way that the effective magnetic field
penetration depth $\Lambda = \lambda^2 / d$ is large compared to
the disk radius $\Lambda \gg R$. This allows us to take into
account only the external field, i.e., $\mathrm{rot}\mathbf{A} =
\mathbf{B} \equiv \mathbf{H}$. In addition, due to smallness of the
inelastic electron--electron scattering time $\tau_{ee}$ with respect to
the electron--phonon one in experimentally relevant setups,
we assume that the QP energy distribution in a superconducting disk is the quasiequilibrium
Fermi distribution $f_T(E) = [e^{E/k_B T}+1]^{-1}$, characterized by a certain
(electronic) temperature $T \ll T_{\rm cs}$ which may differ from the
bath temperature $T_{\rm ph}$. 
In this regime, the normal
($\mathcal{G}$) and anomalous ($\mathcal{F}$) quasiclassical
Green's functions obey the Usadel equations~\cite{Usadel-PRL70},
which are valid for all temperatures and for distances exceeding
the mean free path $\ell$. We concentrate on the cylindrically
symmetric case, introduce the coordinates ($r,\,\varphi,\,z$) as
it is shown in Fig.~\ref{Fig0}, and look for the homogeneous solution along
the $z$-axis solutions characterized by a certain integer angular
momentum $L$ (called a vorticity), 
\begin{equation}\label{Delta_eq}
 \Delta(\mathbf{r}) = \Delta_L(r)\, \mathrm{e}^{i L \varphi}\,,
\end{equation}
which describe the axisymmetric multiquantum vortex states with
the vortex core located at the center of the disk, $r=0$. The
vorticity $L$ coincides with the angular momentum of the
anomalous Green's function $\mathcal{F}$. In the standard
trigonometrical parametrization the quasiclassical Green's
functions for $L$-th orbital mode can be encoded in the pairing
angle $\theta_L(r)$ as
$$
\mathcal{G} =\cos\theta_L,\quad \mathcal{F} = \sin\theta_L\,
\mathrm{e}^{i L \varphi},\quad \mathcal{F}^\dag = \sin\theta_L\,
\mathrm{e}^{-i L \varphi}\,.
$$
The pairing angle $\theta_L$ obeys the following equation (see, e.g.,
Ref.~[\onlinecite{Anthore-PRL03}])
\begin{eqnarray}\label{Usadel-radial_eq}
    -\frac{\hbar D}{2} \nabla_r^2 \theta_L
    &+& \left[\, \omega_n + \Gamma_L(r,H)\,\cos\theta_L\,
    \right]\,\sin\theta_L \\
    &=& \Delta_L(r)\, \cos\theta_L\,, \nonumber
\end{eqnarray}
where the inhomogeneous depairing parameter
\begin{equation}\label{depair-par}
    \Gamma_L(r,\,H) = \hbar v_L^2 / 2D
\end{equation}
is expressed through the superfluid velocity $\mathbf{v}_s =
(0,\,v_L,\,0)$
\begin{equation}\label{super_velocity}
    v_L = D \left( \frac{L}{r} - \frac{\pi\,H}{\Phi_0} r \right)
\end{equation}
and depends on the external magnetic field $H$.
Here $\omega_n = \pi T ( 2 n + 1 )$ is the Matsubara frequency at the temperature $T$,
$D = v_{\rm F} l / 3$ is the diffusion coefficient, and
$\Phi_0 = \pi \hbar c/e$ is the flux quantum.
The relevant length
scale in the Usadel equations is given by the superconducting
coherence length $\xi_0 = \sqrt{\hbar D / 2 \Delta_0}$, where
$\Delta_0$ is the SC gap at zero temperature.
Further we treat only positive $\omega_n$ values,
due to the symmetry of Usadel equations and $\mathcal{F}$ being an even function of
$\omega_n$, $\mathcal{F}(\mathbf{r},-\omega_n) = \mathcal{F}(\mathbf{r},\omega_n)$.
The singlet pairing potential $\Delta_L(r)$ in Usadel equations (\ref{Usadel-radial_eq}) is
determined self-consistently by the equation
\begin{equation}\label{self-consis_eq}
 \frac{\Delta_L(r)}{g}
 - 2\pi T \sum_{n\geq0}\sin\theta_L\, = 0\,,
\end{equation}
where the pairing parameter $g$ fixes the bare critical
temperature $T_{\rm cs}$ as 
\begin{multline}\label{interaction_param}
       \frac{1}{g} = \sum_{n=0}^{\Omega_{\rm D}/(2\pi T_{\rm cs})} \frac{1}{n + 1/2}
        \simeq\ln\left[\frac{\Omega_{\rm D}}{2\pi T_{\rm cs}}\right] + 2\ln2 + \gamma\ .
\end{multline}
Here $\Omega_{\rm D}$ is the Debye frequency and $\gamma \simeq
\mathrm{0.5772}$ is the Euler--Mascheroni constant. The equations
(\ref{Usadel-radial_eq}),(\ref{self-consis_eq}) in the disk bulk should be
accompanied by the boundary conditions at the disk edge
$r = R$ for the order parameter $\Delta_L$ and the pairing angle $\theta_L$:
\begin{equation}\label{edge_bound-cond}
 \left.\frac{d \Delta_L} {d r}\right|_{R} = 0\,, \qquad \left.\frac{d \theta_L} {d r}\right|_{R} = 0\,.
\end{equation}
\begin{widetext}
Both the Usadel~(\ref{Usadel-radial_eq}) and the
self-consistency~(\ref{self-consis_eq}) equations can be obtained
by variation of the free energy functional
\begin{equation}\label{usadelfull}
    F_L = 2 \pi N_0 d \left( \pi T \sum\limits_{\omega_n < \Omega_{\rm D}}
        \int\limits_{0}^{R} r\,d r \left\{ \hbar D \left(\frac{\partial\theta_L}{\partial  r}\right)^2
        + 2 \Gamma_L\,\sin^2\theta_L  - 4 \omega_n \cos\theta_L - 4 \Delta_L \sin\theta_L \right\}
        + \frac{1}{ g } \int\limits_{0}^{R} r\,d r\, \Delta_L^2 \right)\,,
\end{equation}
with $N_0$ representing the density of states at the Fermi level for a
spin projection. 

The power absorbed in the disk is associated with the heat
transferred to phonons emitted by thermal QPs. In general
inhomogeneous case, the electron-phonon heat flow $\dot{Q}_L(r)$
across the central part of the disk of the a radius $r \le R$ into
the phonon bath for the orbital mode $L$ is given by the
expressions~\cite{Timofeev-prl09,Maisi-prl13} 
%
\begin{eqnarray}
   \dot{Q}_L(r) = 2 \pi d \int\limits_0^r d r'\, r'\, \mathcal{P}_L(r') \ , \label{Q_eph_gen_result1}  \quad
   \mathcal{P}_L(r) =  \int\limits_{0}^{\infty}\frac{\Sigma \left[ n_T(\epsilon)
            - n_{T_{\rm ph}}(\epsilon)\right]}{24 \zeta(5) k_B^5}\epsilon^3  d\epsilon \int\limits_{-\infty}^{\infty}
            M_L(r,E,\epsilon) \left[ f_T(E) - f_T(E+\epsilon)\right] dE \ .
\end{eqnarray}
%
Here $\zeta(s)$ is the Riemann zeta function, $k_B$ is the
Boltzmann constant, $n_{T}(\epsilon)=[\exp(\epsilon/k_B
T)-1]^{-1}$ is the Bose function with a temperature $T$. The
kernel of the integrals~(\ref{Q_eph_gen_result1})
\begin{eqnarray}\label{M-function}
    M_L(r,\,E,\,\epsilon) = N_L(r,\,E)\,N_L(r,\,E+\epsilon)
            - B_L(r,\,E)\,B_L(r,\,E+\epsilon)\,,
\end{eqnarray}
with the local density of states (LDOS) $N_L(r,\,E)$ and superconducting correlations $B_L(r,\,E)$
\begin{equation}\label{N_L-B_L}
    N_L(r,\,E) = \left.\mathrm{Re} [\,\cos\theta_L(r)\,]\right|_{\omega_n = - i
    E}\,,\quad
    B_L(r,\,E) = \left.\mathrm{Im}[\,\sin\theta_L(r)\,]\right|_{\omega_n = - i E}\,,
\end{equation}
depending on the radial coordinate $r$, due to the spatial dependence
of the normal ($\sim\cos\theta_L$) and anomalous
($\sim\sin\theta_L$) Green's functions in the disk.
\end{widetext}

To calculate $M_L(r,\,E,\,\epsilon)$ we have to solve the Usadel
equation~(\ref{Usadel-radial_eq}) with the inhomogeneous depairing
parameter
$\Gamma_L(r,\,H)$~(\ref{depair-par}),~(\ref{super_velocity}) for
$\omega_n = - i E$ in the presence of the $L-$quantum vortex in
the disk center. The solution of the Usadel equation gives us the
LDOS $N_L(r,\,E)$ and the spectral gap value $E_{\rm g}$ governing
the contribution to the thermal relaxation mechanisms (see, e.g.,~\cite{Taupin-Khaymovich-NatCom16,Nakamura17}). From the
experimental point of view, LDOS $N_L(r,\,E)$ can be measured
directly  via the local differential conductance (e.g., in the
scanning tunneling spectroscopy setting):
\begin{eqnarray}
 G_L(r,\,V) &=& \frac{d I/d V}{\left(d I/d V\right)_N} \label{loc-diff-cond} \\
 &=& \int\limits_{-\infty}^{\infty} d E\, %
 \frac{N_L(r,\,E)}{N_0}
 \frac{\partial f_T(E - e V)}{\partial V}\,\ . \nonumber
\end{eqnarray}
Here $V$ is the applied bias voltage, $(dI/dV)_N$ is the
conductance of the normal metal junction.

\section{Giant vortex states in mesoscopic disk} \label{GiantVortexStates}

\begin{figure}[t!]
\includegraphics[width=0.4\textwidth]{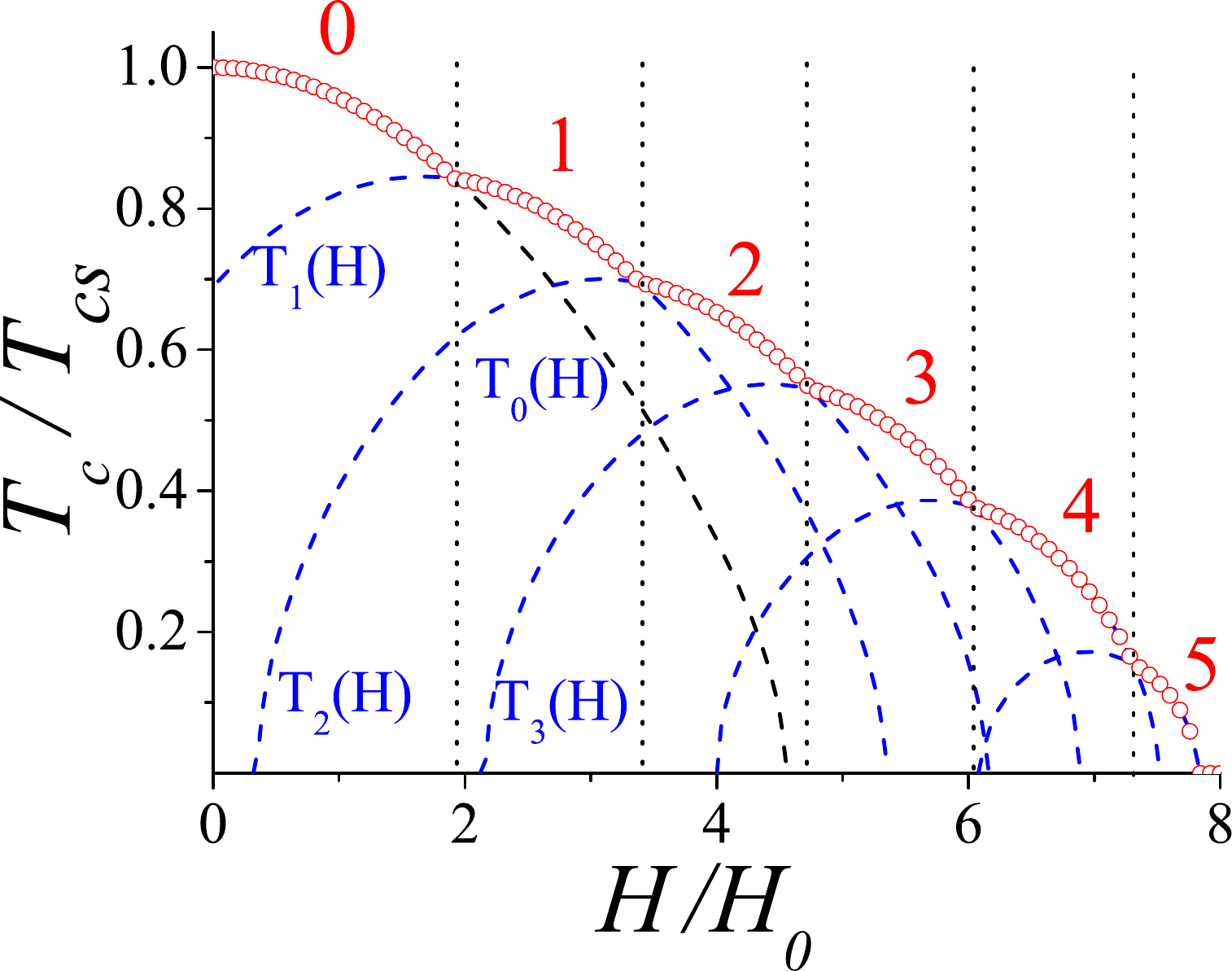}
\caption{(Color online) \textbf{Schematic dependence of the critical
temperature $T_{L}$ ($L = 1 \div 5$) (dashed blue lines) and
$T_{\rm c}$ (red symbols)  on the external magnetic field $H$.}
The vorticity $L$ is denoted by the numbers near the curves.
The values of the magnetic field $H = H_L$ corresponding to the switching of the
orbital modes between $L$ and $L+1$ are shown by the dotted vertical lines.}
\label{Fig2-Tc(H)}
\end{figure}

The phase boundary $T_{\rm c}(H)$ of the mesoscopic SC disk is
known to exhibit an oscillatory behavior similar to the well-known
Little-Parks oscillations~\cite{Little-PRL62,Parks-PRA64}, caused
by the transitions between the states with different angular
momenta $L$. Here $T_{\rm c}(0)=T_{\rm cs}$. Figure~\ref{Fig2-Tc(H)} shows a typical dependence of
the critical temperature
$$
 T_{\rm c}(H) = \underset{L}{\rm max} \left\{\,
 T_{L}(H)\,\right\}\,,
$$
on the external magnetic field $H$, affected by these transitions.
The values of the normalized flux $\phi_L = \pi H_L R^2 / \Phi_0 =
H_L / H_0$ through the disk of a radius $R$, where the switching
of the orbital modes $L \rightleftarrows L+1$ takes place, do not
depend on the disk radius $R$, and can be found using the of
Eqs.~(\ref{Usadel-radial_eq}),~(\ref{self-consis_eq}) linearized
in the anomalous Green's function ($\cos\theta_L \simeq 1$,
$\sin\theta_L \simeq \theta_L$)~\cite{Samokhvalov-PRB19-DOS} or
within the Ginzburg-Landau formalism~\cite{jadallah} based on the free energy~\eqref{usadelfull}.
Note that this energy-based approach~\eqref{usadelfull} does not take into account
possible hysteresis in the orbital-mode switching, caused by the barriers between different free-energy minima.
%
%
%
\begin{figure}[bh!]
\includegraphics[width=0.4\textwidth]{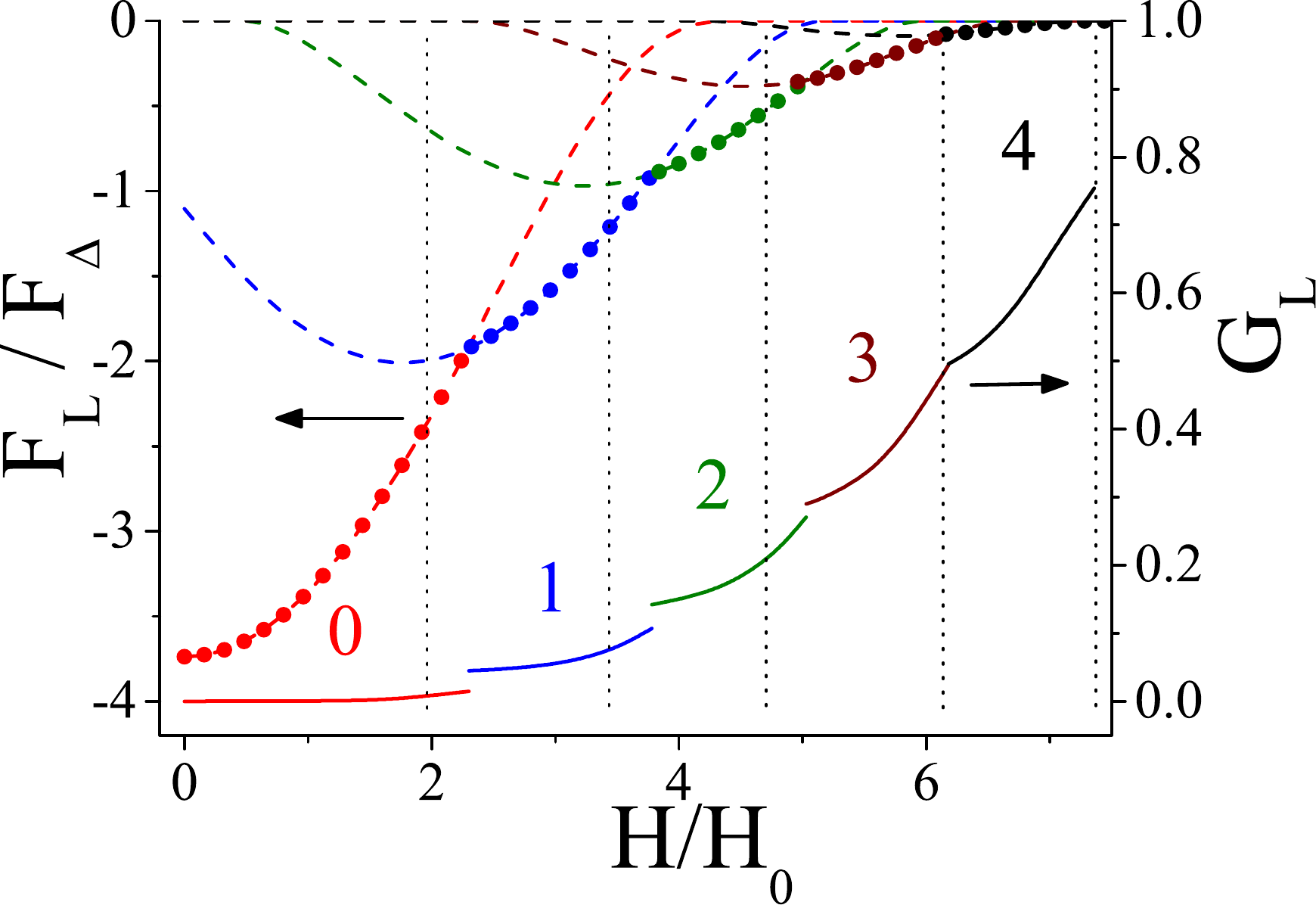}
\caption{(Color online) \textbf{The magnetic-flux $\phi = H /
H_0$-dependence of the normalized zero bias conductance (ZBC)
$G_L(0\,,R)$~(\ref{loc-diff-cond}) at the edge of the SC disk
(solid lines) and the free energy $F_L$~(\ref{usadelfull}) (symbol
$\bullet$)} for the temperatures $T = 0. 1 T_{\rm cs}$. The disk
radius is $R = 4 \xi_0$, the SC coupling constant is $g = 0.18$
The dependence $F_L(\phi)$ for fixed vorticity $\mathrm{L=0 \div
4}$ are shown by the dashed lines. Like in Fig.~\ref{Fig2-Tc(H)},
the corresponding values of vorticity $L$ are denoted by the
numbers near the curves, while the values of the flux $H_L / H_0 =
\phi_L$ corresponding to the switching of the orbital modes
between $L$ and $L+1$, with different critical temperatures
$T_{\rm c}$, are shown by the dotted vertical lines. Here the free
energy is normalized by $F_\Delta = \pi \hbar D N_0 d \Delta_0$.}
\label{Fig3-FGL(H)}
\end{figure}
%

%
\begin{figure*}[t]
\includegraphics[width=0.22\textwidth]{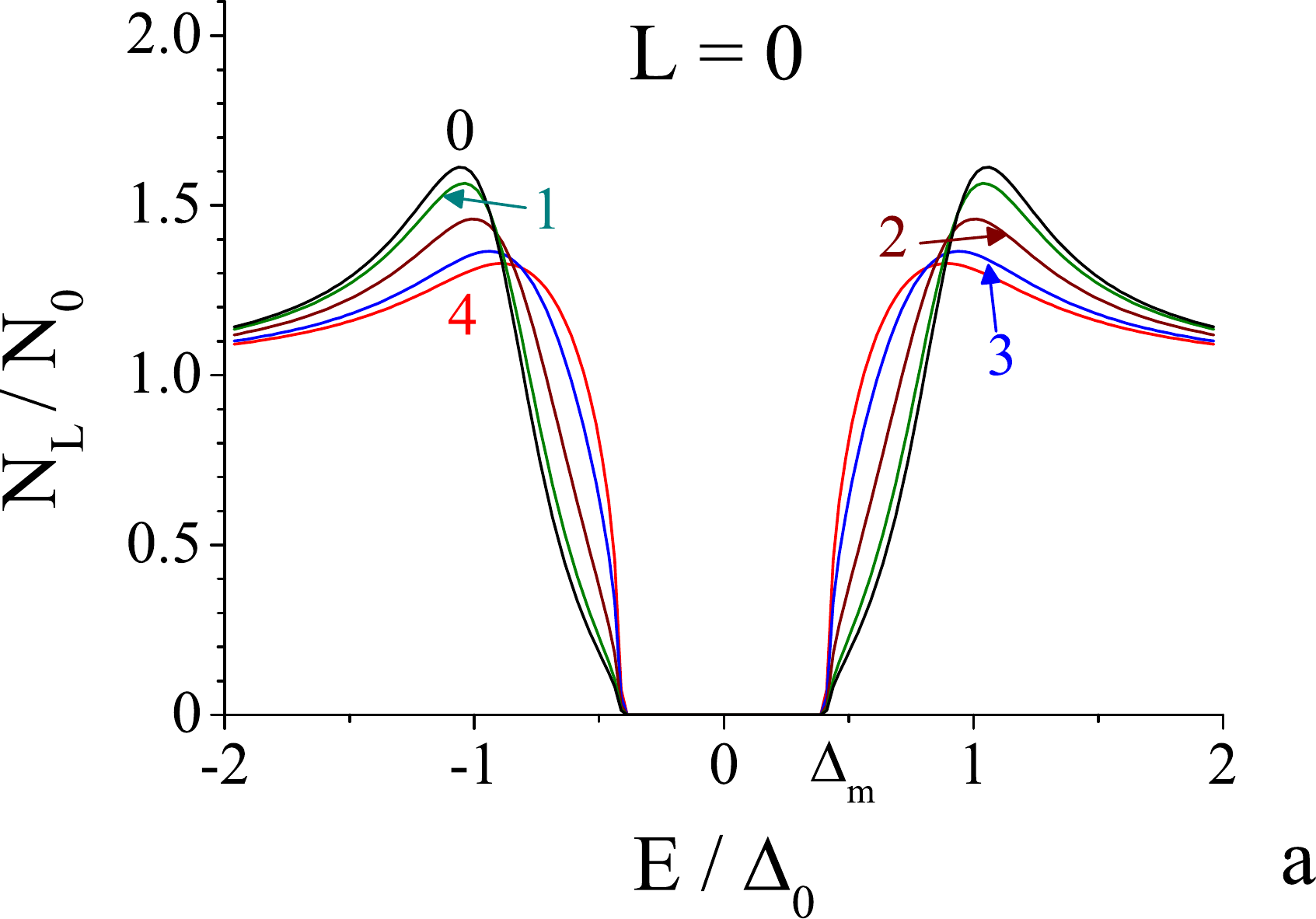}
\includegraphics[width=0.22\textwidth]{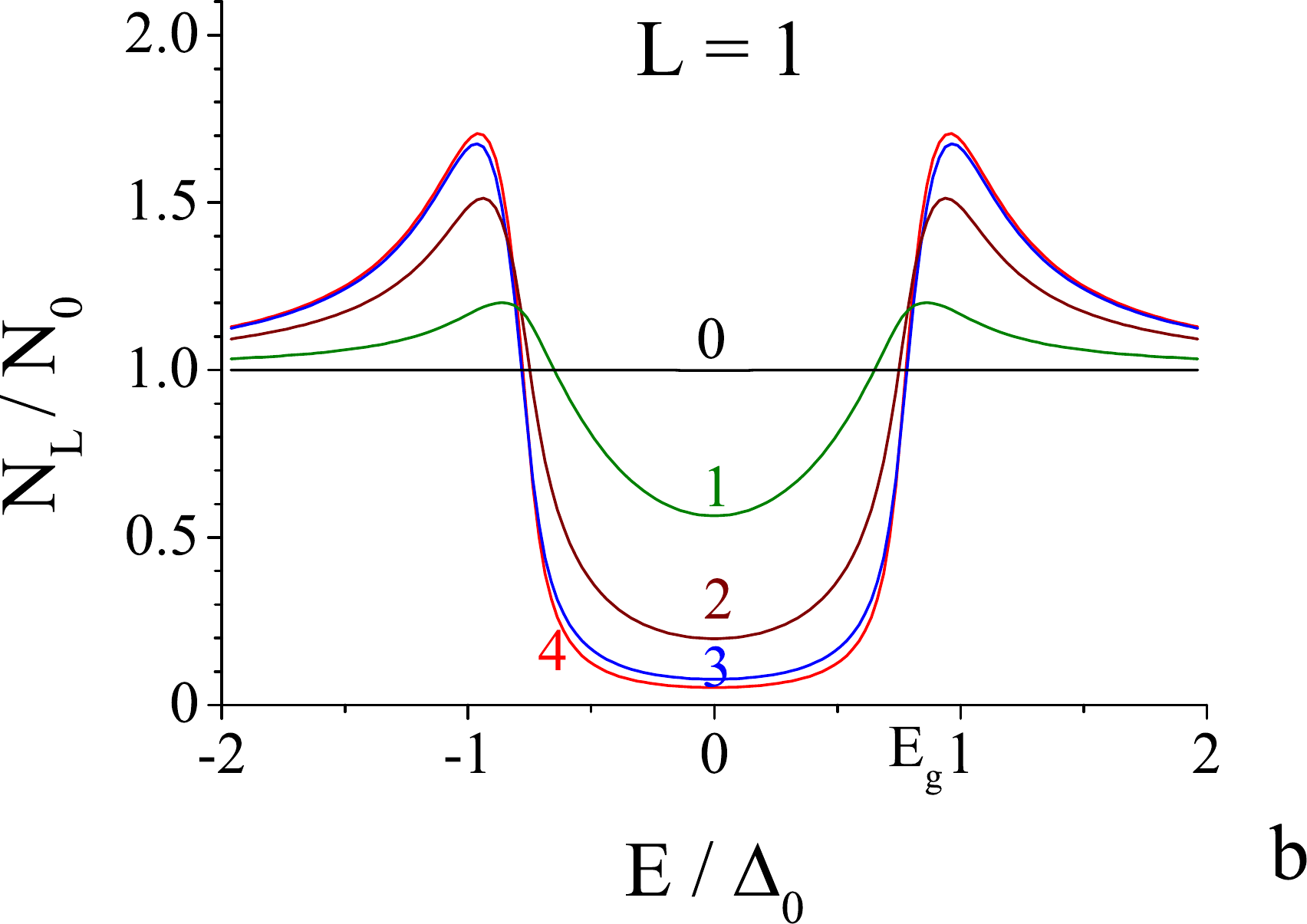}
\includegraphics[width=0.22\textwidth]{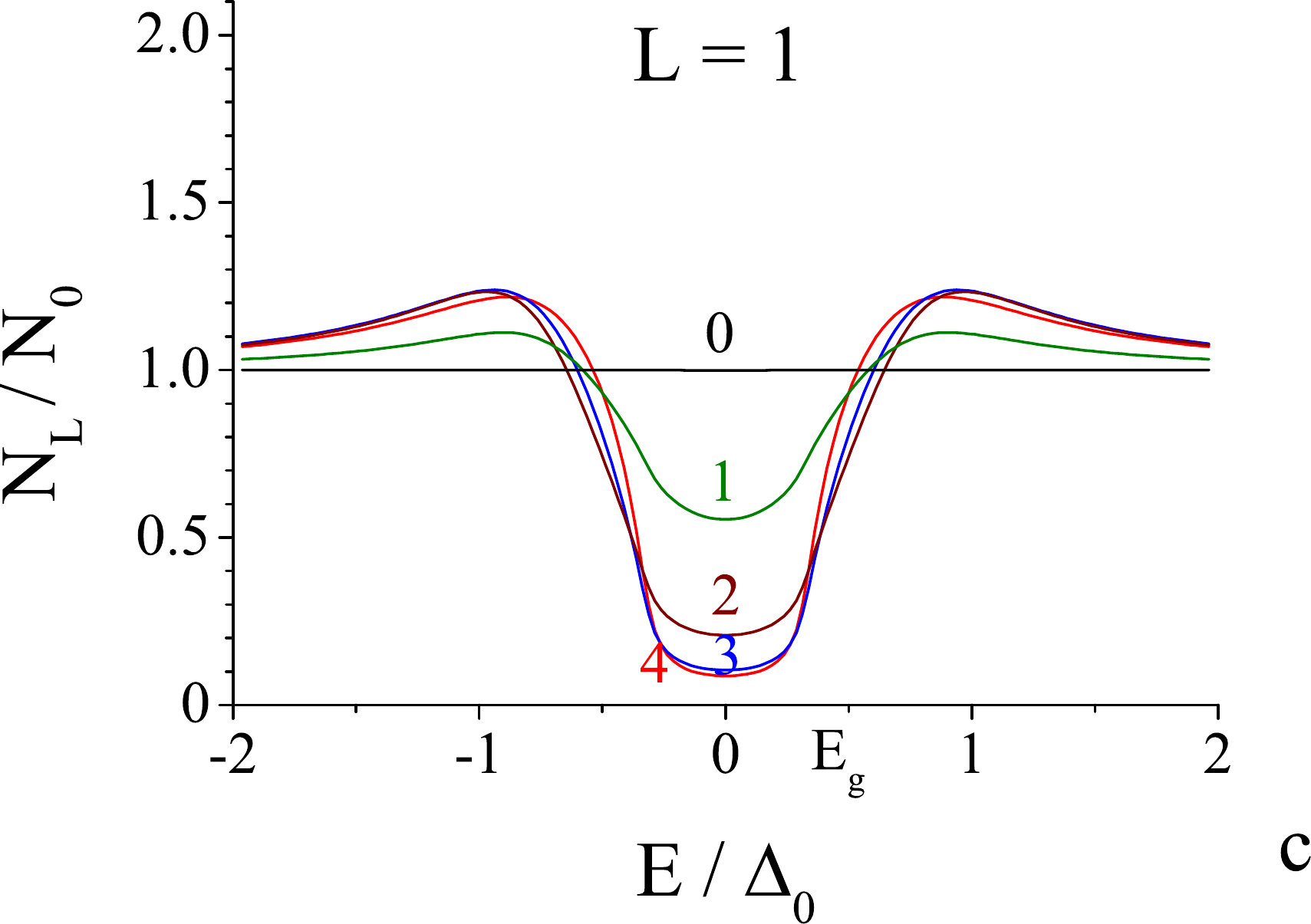}
\includegraphics[width=0.22\textwidth]{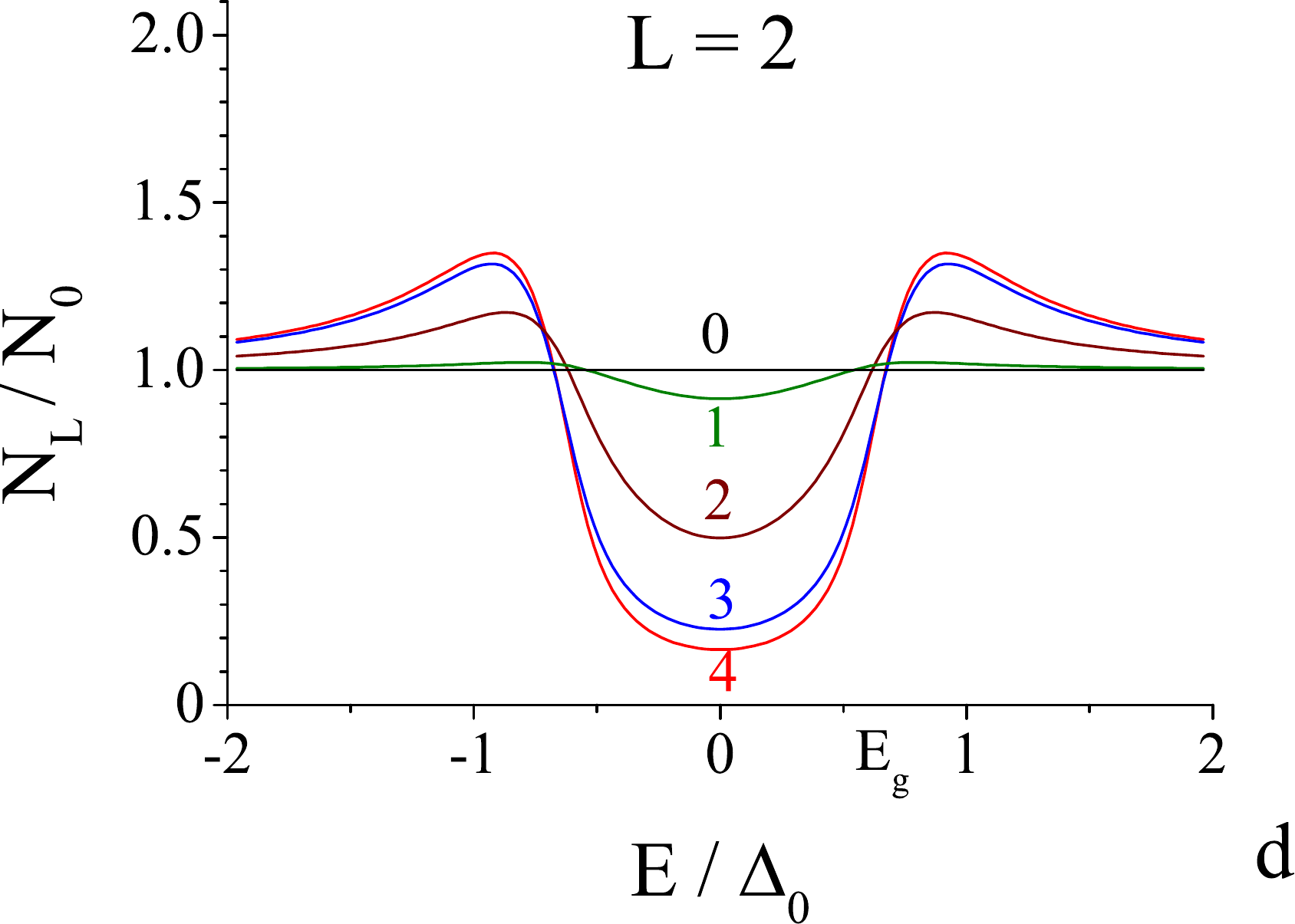}
\includegraphics[width=0.22\textwidth]{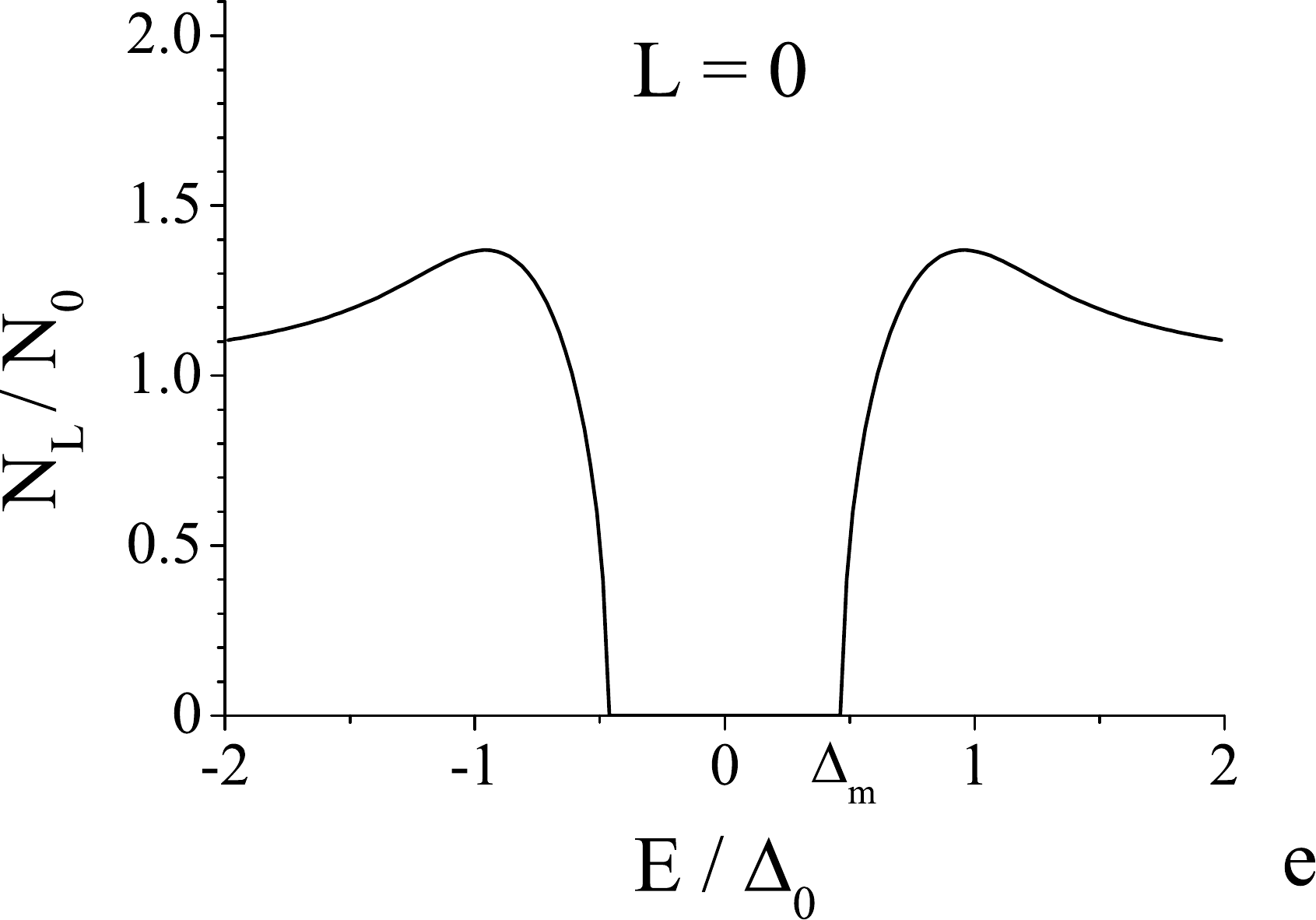}
\includegraphics[width=0.22\textwidth]{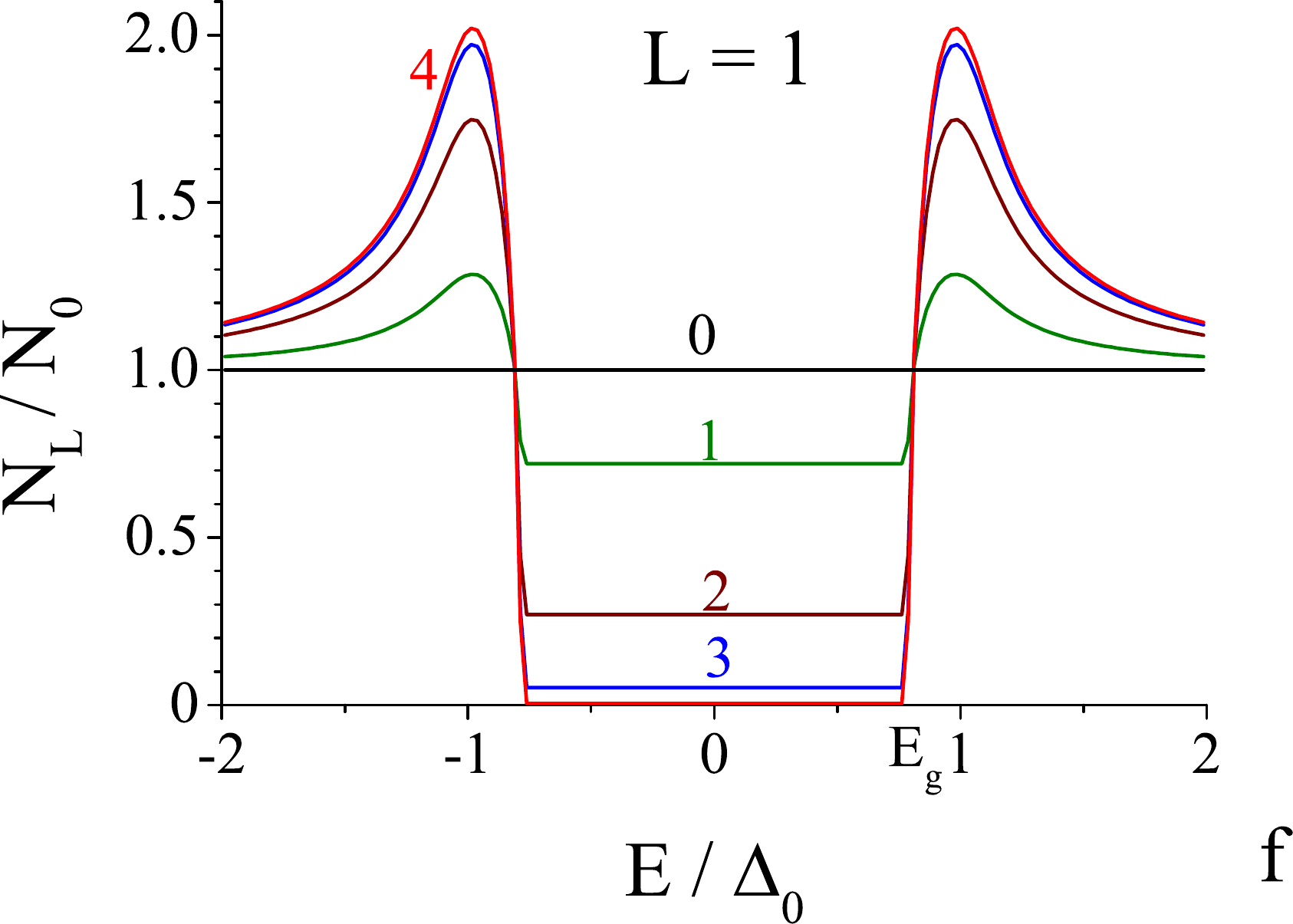}
\includegraphics[width=0.22\textwidth]{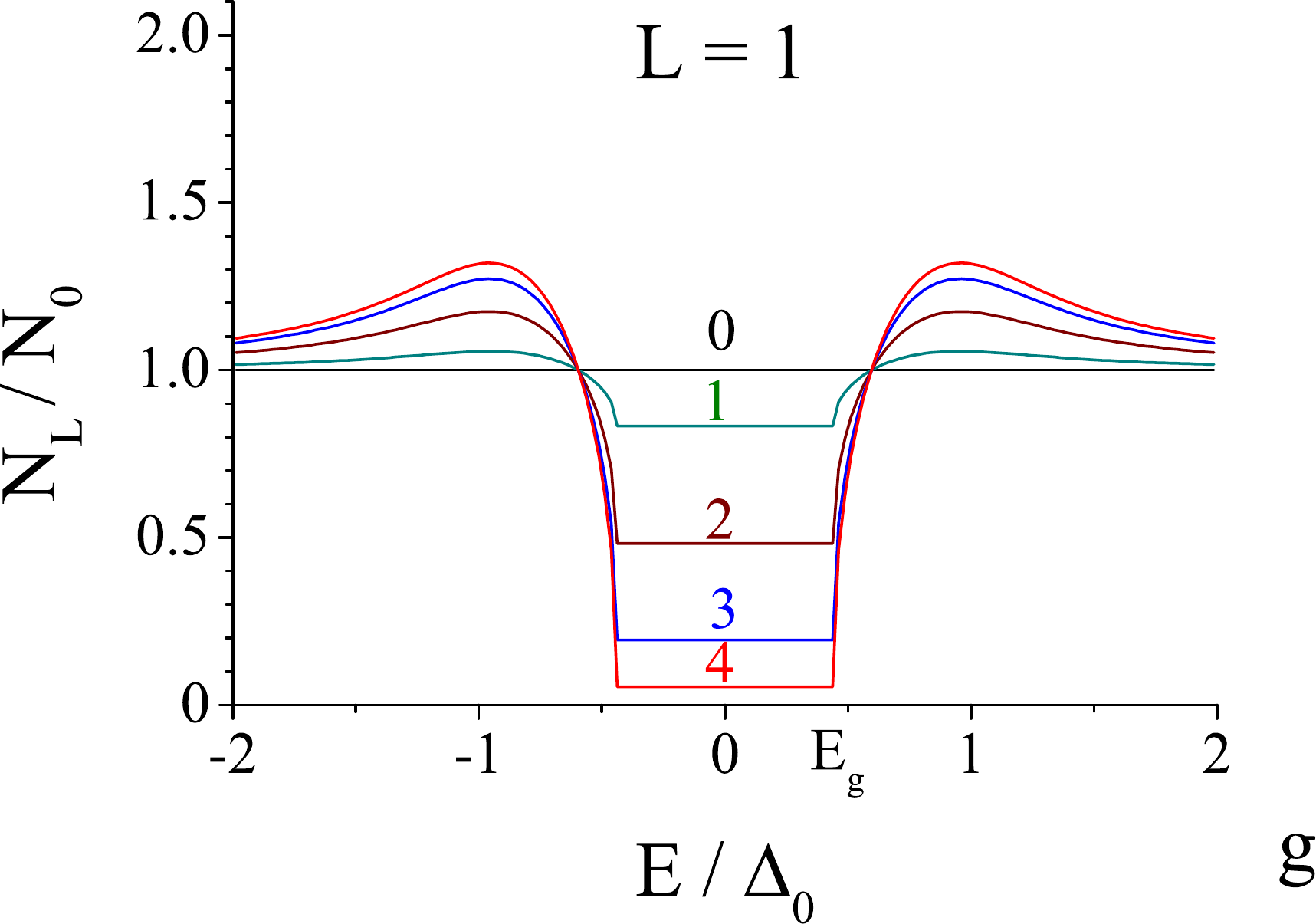}
\includegraphics[width=0.22\textwidth]{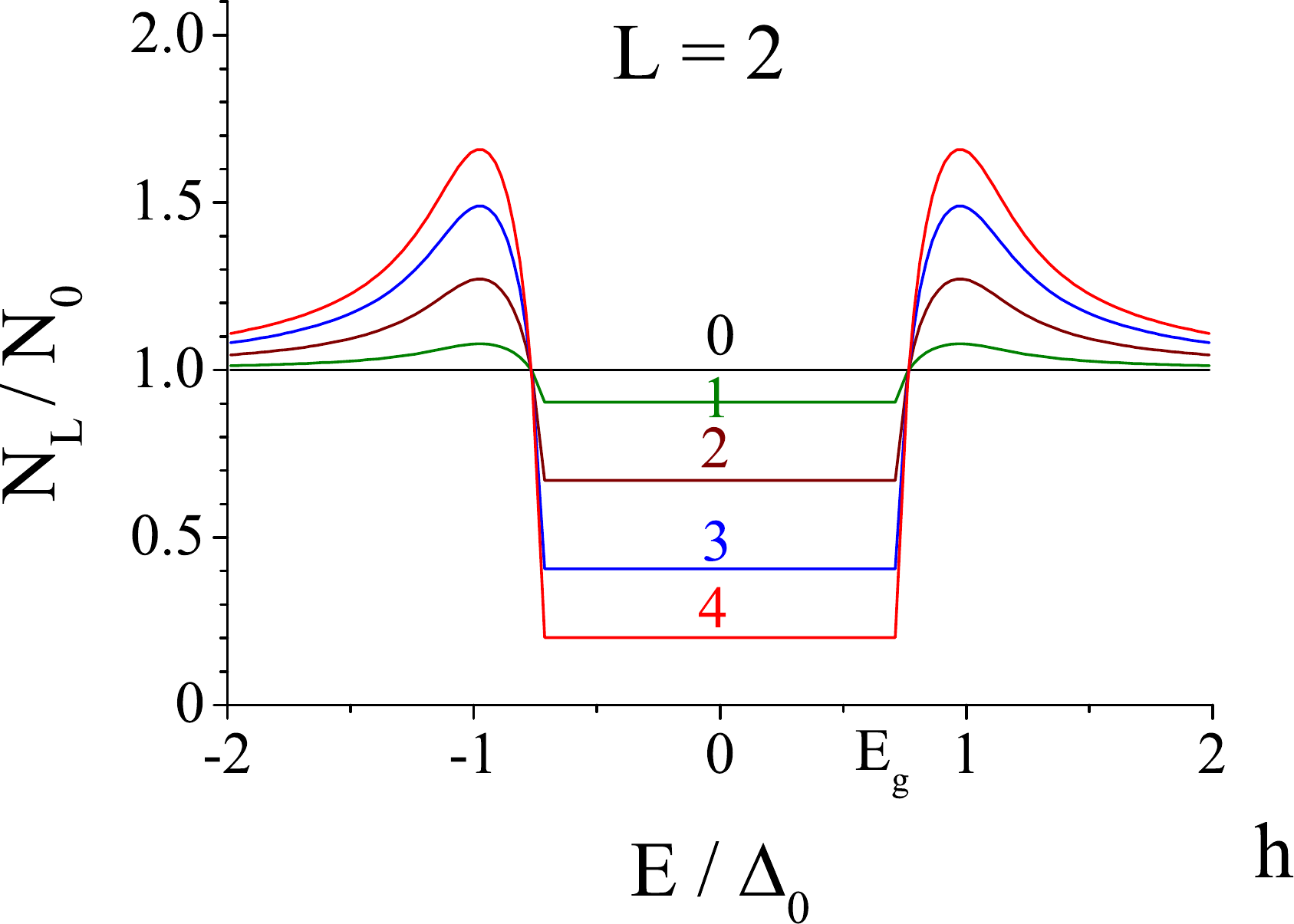}
\caption{(Color online) \textbf{Comparison of the
exact~(\ref{N_L-B_L})} (panels a-d) \textbf{and
approximate~(\ref{N_L-appr})} (panels e-h) \textbf{LDOS
$N_L(r\,,E)$ versus energy $E$ for several distances $r$ from the
vortex core, vorticities $L$, and magnetic fields $H$.}
The SC disk radius $R = 4 \xi_0$ and the temperature $T = 0.1
T_{\rm cs}$ are fixed for all panels. The magnetic field values
$H$ are: (a, b, e, f)~$H = H_{s1} = 2.24\,H_0$; (c, d, g, h)~$H =
H_{s2} = 3.84\,H_0$. The distances $r$ (in the units of $\xi_0$)
are shown by numbers of the corresponding colors near the curves.
The soft minigap at the edge $E_\mathrm{g}$ in (b-d) is assumed to
correspond to the maximal slope of the energy dependence of the
LDOS $d N_L(R,E) / d E$.} \label{Fig-DOS}
\end{figure*}
%

In order to analyze the transitions between different vortex
states far from the phase transition line $T \ll T_{\rm c}(H)$ we
need to use the nonlinear Usadel theory~(\ref{Usadel-radial_eq}~-~\ref{edge_bound-cond}). The Usadel
equations have been solved numerically for different vorticities
$L$ which allowed us to calculate and compare the values of the
free energy $F_L$~(\ref{usadelfull}). Figure~\ref{Fig3-FGL(H)}
shows the magnetic field dependence of the free energy~(\ref{usadelfull}) and
the zero bias conductance $G_L(R,\,0)$~(\ref{loc-diff-cond}) at the Fermi level for a small disk radius
$R = 4 \xi_0$ and the temperature $T = 0. 1 T_{\rm cs}$. The
curves illustrate the switching between the states with different
vorticities $L = 0 \div 4$, which is similar to the Little-Parks-like
switching of the critical temperature $T_{\rm c}(H)$ shown in Fig.~\ref{Fig2-Tc(H)}.

Sequential entries of vortices produce a set of branches $F_L$
with different vorticity $L$ on the $F(H)$ and $dI/dV(H)$ curves.
The transitions between different vortex states are accompanied by
an abrupt change in the zero bias conductance (ZBC) at the disk
edge, which is attributed to the entry/exit of a vortex while
sweeping the magnetic field. Note, that the field values $H_{sL}$
at which the jumps in vorticity ($L-1 \to L$) occur at low
temperature are always larger than the values $H_L$ found from the
calculations of the critical temperature behavior $T_{\rm c}(H)$.
For a fixed disk radius $R$ the direction of jumps
(upward/downward) in the dependence of ZBC $G_L(R,\,0)$ vs
magnetic field depends on the temperature $T$ and reflects a
crossover between the edge-dominated and core-dominated regimes in
the magnetic field dependence of the tunneling conductance (see
Ref.~\onlinecite{Samokhvalov-PRB19-DOS} for details).
Figure~\ref{Fig-DOS} illustrates evolution of the spatially
resolved LDOS $N_L(r\,,E)$ vs energy $E$ for values of the magnetic
field $H = H_{s1}\,,H_{s2}$ corresponding to the vorticity
switching. In the Meissner state ($L = 0$), the hard minigap
$\Delta_{\rm m}$ in the spectrum survives and $N_{L=0}(r,\,E <
\Delta_{\rm m}) = 0$ until the first vortex entry,
Fig.~\ref{Fig-DOS}(a). In vortex states ($L \ge 1$),
Figs.~\ref{Fig-DOS}(b-d), the density of states $N_L(0\,,E)$ is
equal to the normal-metal electronic density of states at the
Fermi level $N_0$, indicating a full suppression of the spectral
gap in the disk center due to the vortex entry. At the same time,
at the edge of the disk the superconductivity survives though the
gap becomes soft, $0 < N_L(R\,, E) < N_0$. The soft minigap at the
edge $E_\mathrm{g}$ is assumed to correspond to the maximal slope
of the energy dependence of the LDOS $d N_L(R,E) / d E$.
Figures~\ref{Fig-DOS}(a-b) illustrate the switching between the
states with hard and soft gaps at $H = H_{s1} \simeq 2.24\,H_0$
when the Meissner state ($L = 0$) transforms to the single--vortex
state ($L=1$). At fixed vorticity $L=1$ as the magnetic field
increases, the LDOS $N_L(r\,,E)$ smoothly evolves with the
simultaneous decrease of the soft minigap at the disk edge $E_{\rm
g}$, Figs.~\ref{Fig-DOS}(b-c).
This smooth evolution of the LDOS is interrupted at $H = H_{s2}
\simeq 3.84\,H_0$ by a next vortex entry which restores the
superconductivity near the disk edge and results in increase of
the soft minigap $E_{\rm g}$, Figs.~\ref{Fig-DOS}(c-d).
Certainly, these drastic changes in the LDOS $N_L(r,\,E)$ and the
spectral gap value $E_{\rm g}$ directly impact the thermal
relaxation mechanisms and manifest themselves in peculiarities of
the magnetic field dependence of the electron-phonon heat flow
$\dot{Q}$, which we consider further.
%
\begin{figure}[t]
\includegraphics[width=0.47\textwidth]{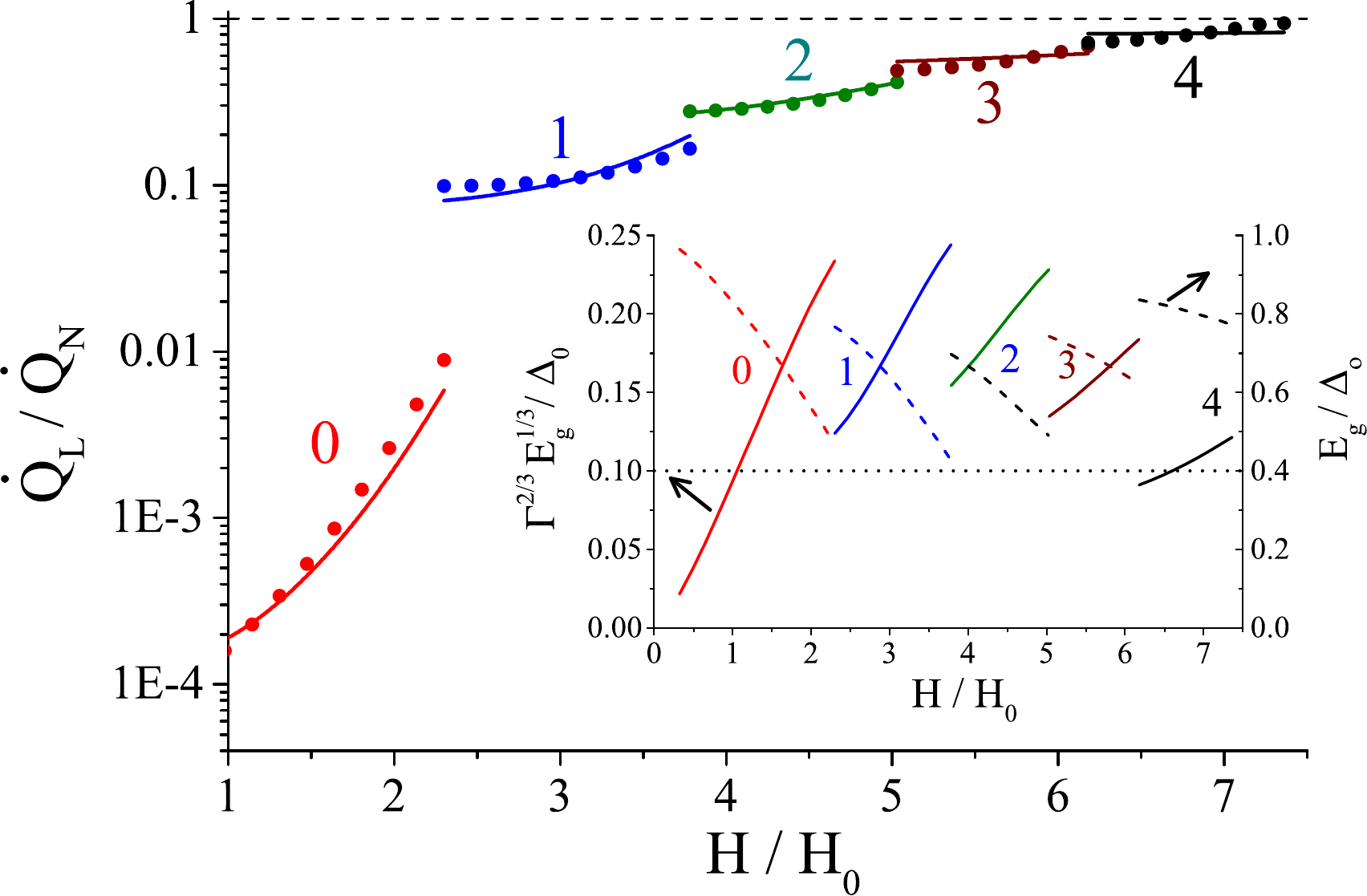}
\caption{(Color online) \textbf{The dependence of the
electron--phonon heat flow $\dot{Q}_L(R)$~(\ref{Q_eph_gen_result1}) 
on the magnetic flux  $\phi = H / H_0$ across the SC disk} of the
radius $R = 4 \xi_0$ for the temperatures $T = 0. 1 T_{\rm cs}$
(filled circles). The solid lines show the results of fitting
$\dot{\bar{Q}}_L$~(\ref{Q_eph_gen_result1-Appr}),~(\ref{Q_eph_gen_result2-Appr})
in homogeneous approximation described in
Sec.~\ref{HeatFluxHomoSect}. The inset shows the dependence of the
hard gap $\bar{E}_g$ (solid lines) and the value
$\bar{\Gamma}^{2/3} \bar{E}_g^{1/3}$ (dashed lines) determined by
the relations~(\ref{Eg-Delta}) on the magnetic flux $\phi$. The
dotted line shows the value of the temperature $T$. The numbers
near the curves denote the corresponding values of vorticity $L =
0 \div 4$. ($\dot{Q}_N = \Sigma \mathcal{V} T^5$, $\mathcal{V} =
\pi R^2 d$).} \label{Fig4-Q(H)}
\end{figure}
%

\section{Electron-phonon heat flow in giant vortex configurations} \label{E-PhHeatFlux}

As a next step we calculate the thermal relaxation rate in the
superconducting disk assuming the electronic temperature $T$ to be
much larger than the bath (phonon) temperature $T_{\rm ph}$, thus,
neglecting for simplicity all the exponential terms like
$e^{-E_g/k_B T_{\rm ph}}$. Filled circles in Fig.~\ref{Fig4-Q(H)}
show the magnetic field dependence of the total electron--phonon
heat flow $\dot{Q}$ versus the applied magnetic field $H$ for a
small disk. Here and further we focus on the disk radius $R = 4
\xi_0$ and the temperature $T = 0.1 T_{\rm cs}$ (if not mentioned
otherwise). The total $\dot{Q}(R)$ curve consists of several
separate branches $\dot{Q}_L$~(\ref{Q_eph_gen_result1})
corresponding to the states with different vorticity $L = 0 \div
4$. The transitions between different vortex states are visualized
by abrupt changes (or jumps) $\triangle\dot{Q}$ in the heat flow
at $H = H_{sL}$, where switching of the orbital modes $L - 1
\rightleftarrows L$ takes place. The smooth growth of the
electron--phonon heat flow $\dot{Q}_L$ occurs while sweeping the
magnetic field up within the branch $L$, $H_{sL} < H <
H_{s(L+1)}$, due to decrease of the value of the spectral gap
(soft or hard) and increase in subgap LDOS at the disk
edge~\cite{Samokhvalov-PRB19-DOS}. The especially strong growth of
$\dot{Q}_L$ appears in the Meissner state ($\dot{Q}_0(H_{s1}) /
\dot{Q}_0(0) \sim 10^2$) owing to essential suppression of the
hard minigap $\Delta_m$ in the spectrum by the screening currents,
Fig.~\ref{Fig-DOS}(a).

\begin{figure*}[t!]
\center{
\includegraphics[width=0.333\textwidth]{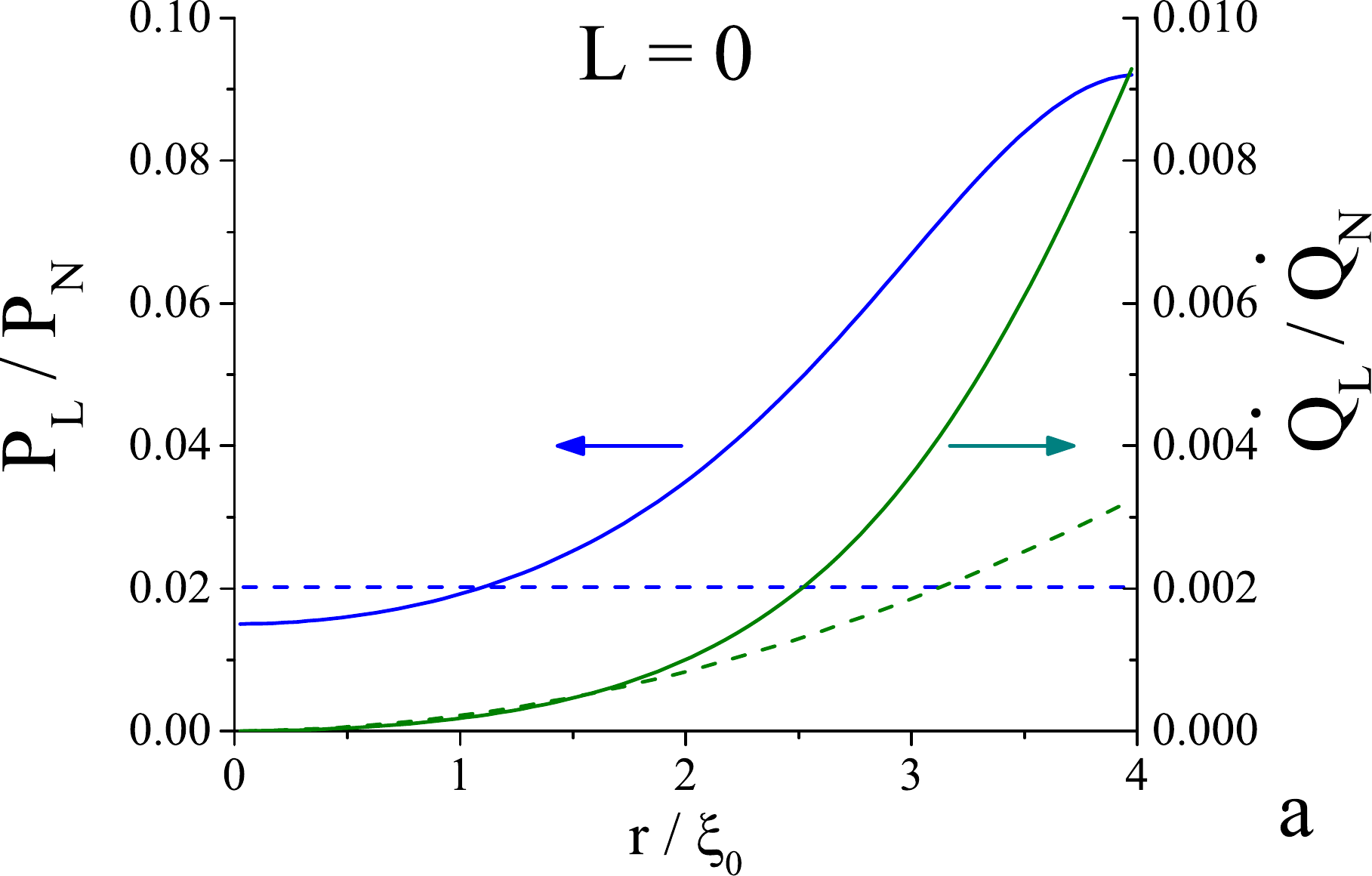}
\includegraphics[width=0.333\textwidth]{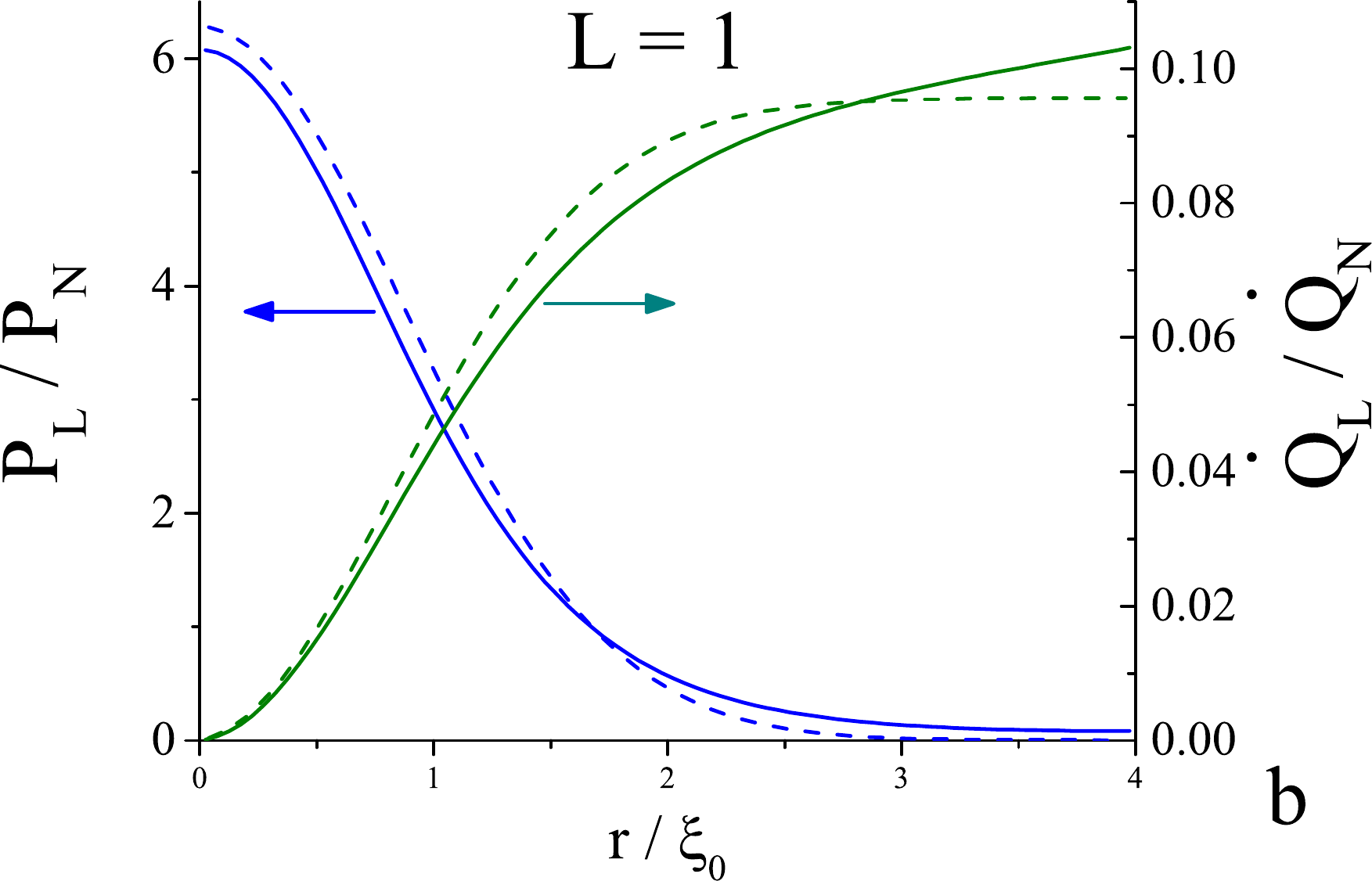}\\
\includegraphics[width=0.333\textwidth]{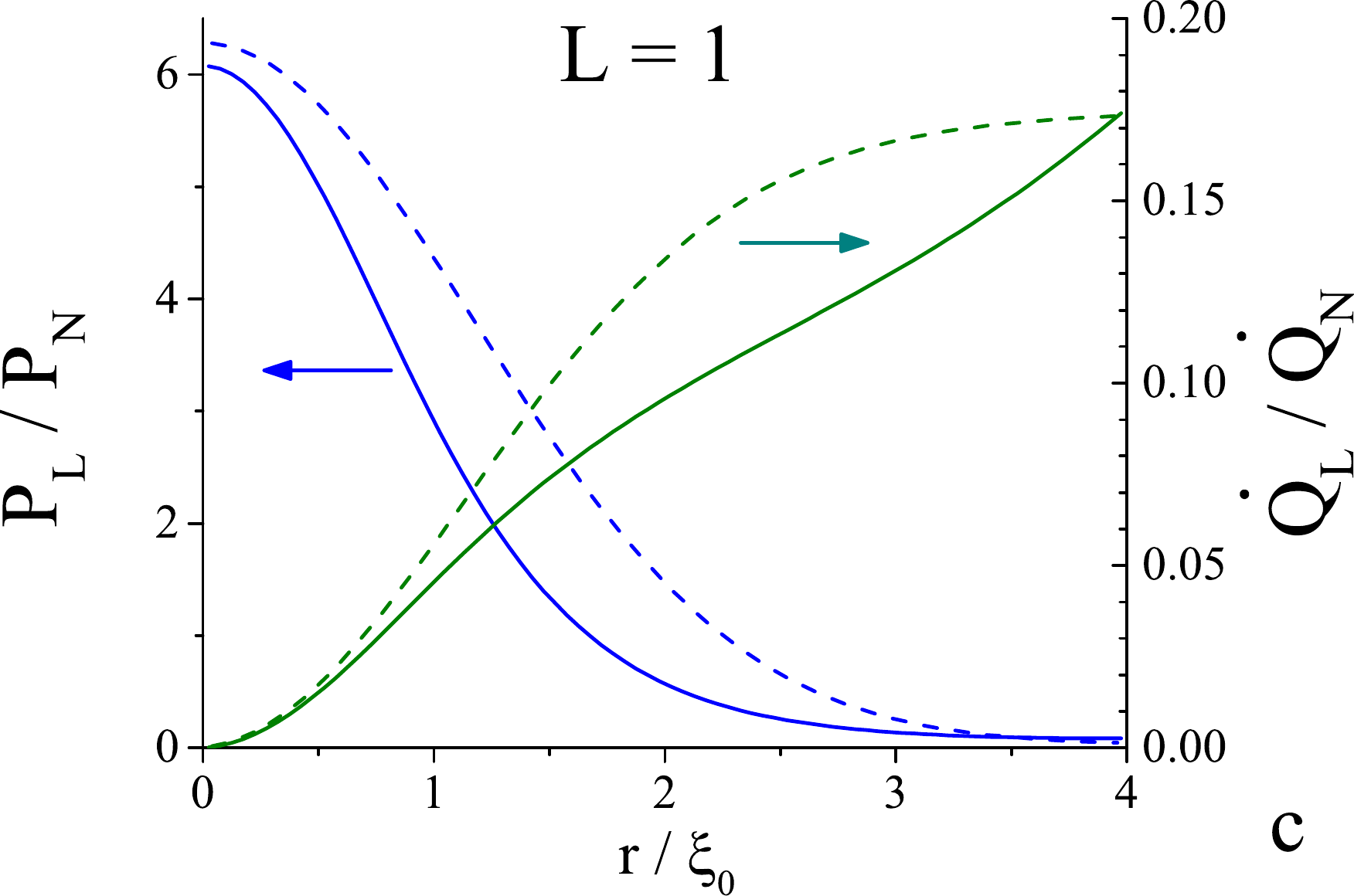}
\includegraphics[width=0.333\textwidth]{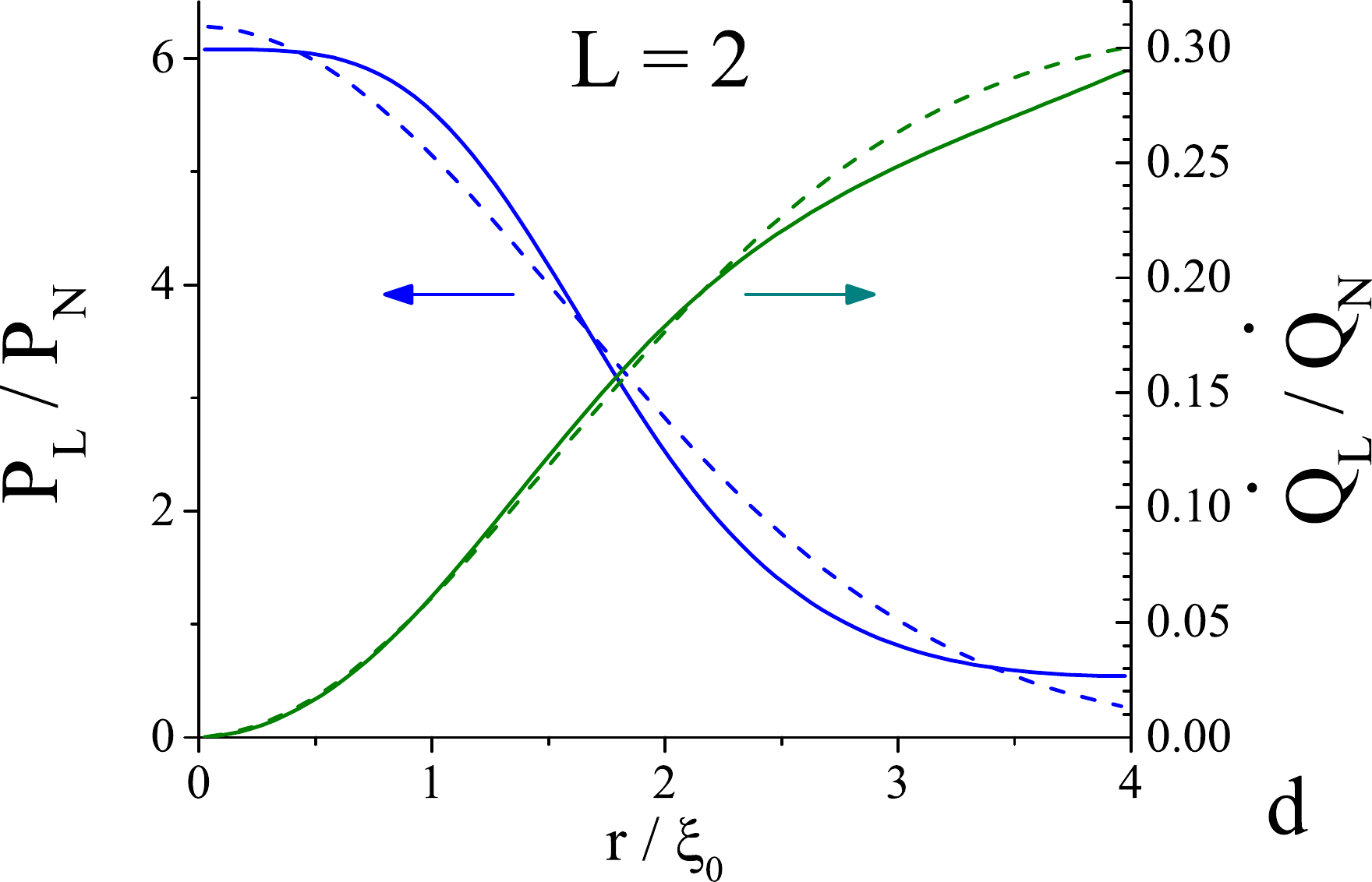}
\caption{(Color online) \textbf{The spatially resolved heat flow
$\dot{Q}_L(r)$ and the density of the electron--phonon heat flow
$P_L(r)$~(\ref{Q_eph_gen_result1}) (solid lines) for several
values of magnetic field $H$}: (a,~b)~$H = H_{s1} = 2.24\,H_0$;
(c,~d)~$H = H_{s2} = 3.84\,H_0$. The dashed lines show the results
of fitting
$\dot{Q}_L(r)$~(\ref{Q_eph_gen_result1-Appr}),~(\ref{Q_eph_gen_result2-Appr})
for $Z_{\mathcal{D}_L}(r) = \mathrm{e}^{-r^2 / \mathcal{D}_L^2}$
($L \ge 1$) described
in section~\ref{HeatFluxHomoSect}: $D_1 = \mathrm{1.6} \xi_0$, $D_2 = \mathrm{2.7} \xi_0$. %
($R = 4 \xi_0$, $T = 0. 1 T_{\rm cs}$, $\dot{Q}_N = \Sigma
\mathcal{V} T^5$, $P_N = \Sigma T^5$, $\mathcal{V} = \pi R^2 d$).}
}
\label{Fig-PQ}
\end{figure*}

To compare the contributions to the total electron--phonon heat
flow $\dot{Q}(R)$~(\ref{Q_eph_gen_result1}) from the vortex core
and the region near the sample edge with the flowing Meissner
current, we present in Fig.~\ref{Fig-PQ} radial distributions of
the heat flow $\dot{Q}_L(r)$ and the flow density
$\mathcal{P}_L(r)$~(\ref{Q_eph_gen_result1}) for two values of the
applied magnetic field $H = H_{s1}\,,H_{s2}$ corresponding to the
switching between the states $L=0 \to 1$ and $L=1 \to 2$,
respectively. In the Meissner state the spatially resolved heat
flow density $\mathcal{P}_L(r)$ and the heat flow $\dot{Q}_L(r)$
are more pronounced near the disk edge where the screening
superconducting currents have higher density and the SC order
parameter is suppressed, Fig.~\ref{Fig-PQ}(a). A vortex in the
mesoscopic SC disk leads to QP redistribution so that the heat
flow density in the vortex core is always higher than near the
disk edge ($\mathcal{P}_{L \ne 0}(0) > \mathcal{P}_{L \ne 0}(R)$).
Despite this, the contribution of the region outside the vortex
core to the total electron--phonon heat flow appears to be
comparable to the contribution of the vortex core itself because
of the relatively small volume of the core region. Note, that the
significant growth of $\dot{Q}_L(r)$~(\ref{Q_eph_gen_result1}) in
Figs.~\ref{Fig-PQ}(b-d) outside the vortex core region illustrate
the existence of a noticeable interplay between the contributions
from  the subgap states, located in the vortex cores and in the
region with the reduced spectral gap $E_{\rm g}$ near the sample
edge. The
\onecolumngrid
\newpage
\twocolumngrid
\noindent
interplay increases when the field is swept from
$H_{sL}$ to $H_{s(L+1)}$ for a fixed vorticity $L$ and with the
increase in the number of vortices $L$ trapped in the center of
the sample.

The electronic properties of the vortex states will be surely
modified if we further increase the radius of the disk $R$
compared to the coherence length $\xi_0$. In this case the core of
a multiquantum vortex does not extend to the edge of the disk, and
quasiparticles in the vortex core remain well localized near the
disk center. Clearly, in this case the profiles of the heat flow
density $\mathcal{P}_L(r)$ and the heat
flow $\dot{Q}_L(r)$~(\ref{Q_eph_gen_result1}) have to reveal a
local minimum and a plateau, respectively.
%

\begin{figure}[t!]
\includegraphics[width=0.7\columnwidth]{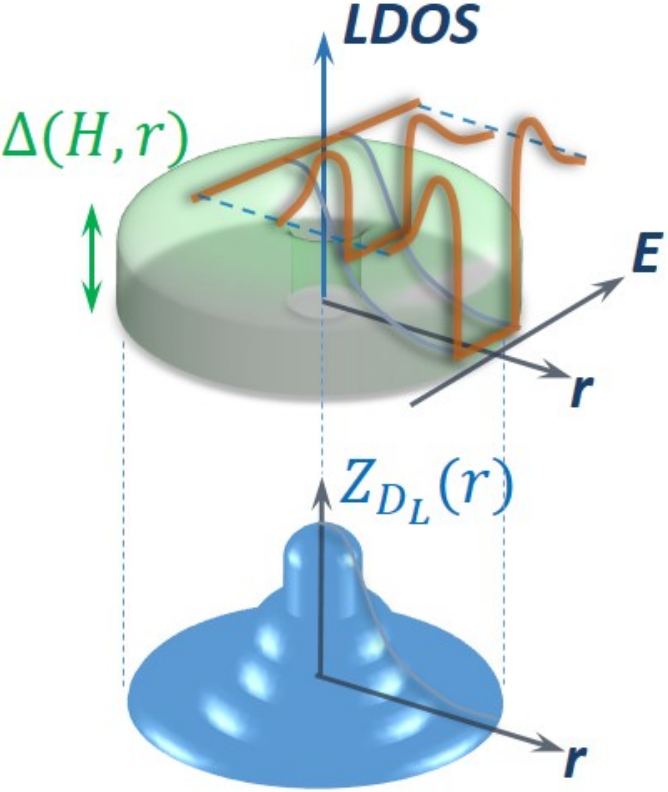}
\caption{(Color online) \textbf{Sketch of the homogeneous
approximation.} (top)~The soft superconducting gap (shown by
semi-transparent green color) in the superconducting disk of the
radius $R$ with the giant $L$-quantum vortex is accompanied by the
energy-dependence of the local density of states (orange lines),
Eq.~\eqref{N_L-appr}, being a superposition of the vortex
$\bar{N}_V(E)=1$ and homogeneous $\bar{N}_L(E)$ LDOS with the
radial profile $Z_{\mathcal{D}_L}(r)$ (shown by two parallel blue
curves). (bottom)~The cylindrical-symmetric representation of the
radial function $Z_{\mathcal{D}_L}(r)$ in the SC disk. }
\label{Fig_homo_cartoon}
\end{figure}

\section{Semiquantitative
homogeneous approximation for the electron--phonon heat flow}\label{HeatFluxHomoSect}

For the analysis of the experimental data it is often useful and
convenient to built some semiquantitative approximations for the
measurable quantities which would allow to avoid the extensive use
of the numerical simulations. To develop such a description  we
try to fit the above numerical simulations of the magnetic field
dependence of the electron--phonon heat flow $\dot{Q}_L(H)$ (see
Fig.~\ref{Fig4-Q(H)}) by a simplified model taking into account
that the overall spectral characteristics and local density of
states of the mesoscopic sample result from the interplay of the
subgap states, located in the vortex core and in the edge regions
with the spectral gap reduced by Meissner currents~\cite{Samokhvalov-PRB19-DOS}. We assume that both the order
parameter $\Delta_L(r)$ and $\theta_L(r)$ function vanish inside
the vortex core and
separate the energy and the coordinate dependence
in LDOS $N_L(r,\, E)$ and superconducting correlations
$B_L(r,\, E)$~(\ref{N_L-B_L}) by the following approximate expressions (see Fig.~\ref{Fig_homo_cartoon})
\begin{eqnarray}
    N_L(r,\, E) &\simeq& \left[ 1 - Z_{\mathcal{D}_L}(r) \right] \bar{N}_L(E)
        + Z_{\mathcal{D}_L}(r) \bar{N}_V(E)\,, \label{N_L-appr} \\
    B_L(r,\,E) &\simeq&  \left[ 1 - Z_{\mathcal{D}_L}(r) \right] \bar{B}_L(E)
        + Z_{\mathcal{D}_L}(r) \bar{B}_V(E)\,.  \label{B_L-appr}
\end{eqnarray}
Here the functions $\bar{N}_V(E) = \mathrm{Re}[\cos\theta_V] = 1$
and $\bar{B}_V(E)= \mathrm{Im}[\sin\theta_V] = 0$ ($\theta_V =
0$), describe normal metal properties of the vortex core region.
A radial profile $Z_{\mathcal{D}_L}(r)$ which describes a
contribution of the subgap states in the vortex core
($Z_{\mathcal{D}_L}(0) = 1$, $Z_{\mathcal{D}_L}(R) \ll 1$)
is assumed to be a monotonically decaying function
with a certain $L$-dependent length scale $\mathcal{D}_L$, see Fig.~\ref{Fig_homo_cartoon}~(bottom).
A residual small contribution of the subgap states related to the vortex core
at the disk edge ($\sim Z_{\mathcal{D}_L}(R)$) corresponds to a
nonzero density of states at the Fermi level and is responsible
for the destruction of the hard gap in the density of states. In
the Meissner state we get $Z_0(r) = 0$ and, thus, $\mathcal{D}_0 = 0$.

The functions $\bar{N}_L(E)$ and $\bar{B}_L(E)$ describe the
averaged characteristics of the superconducting state in the
vortex free phase and outside the vortex core region. To get these
functions we average the Eq.~(\ref{Usadel-radial_eq}) over the
radial coordinate neglecting the gradient terms, i.e. assume the
$\theta_L(r)$ function to depend slowly on $r$. The spatially
averaged value $\bar{\theta}_L$ satisfies to the Usadel equation
for the normal ($\cos\bar{\theta}_L$) and anomalous ($-i
\sin\bar{\theta}_L$) Green's functions~\cite{Maki-SC}
\begin{equation}\label{Usadel-aver_eq}
    \left(\,i E - \bar{\Gamma}_L\,\cos\bar{\theta}_L\,
    \right)\,\sin\bar{\theta}_L = \bar{\Delta}_L\, \cos\bar{\theta}_L\,,
\end{equation}
which describes different depairing effects in dirty
superconductors with the effective depairing energy
$\bar{\Gamma}_L$~\cite{Anthore-PRL03,eg1,eg2,eg3}. The solution
$\bar{\theta}_L$ of the algebraic equation~(\ref{Usadel-aver_eq})
at $\omega_n = - i E$ gives us the energy dependence of the
density of states $\bar{N}_L(E) = \mathrm{Re}[\cos\bar{\theta}_L]$
and the function $\bar{B}_L(E) = \mathrm{Im}[\sin\bar{\theta}_L]$.
The hard gap $\bar{E}_\mathrm{g}$ in the density of states
$\bar{N}_L(E)$ and the order parameter $\bar{\Delta}_L$ are
determined by the
relations 
\begin{equation}\label{Eg-Delta}
    \bar{E}_\mathrm{g} = \bar{\Delta}_L \left( 1 - \gamma_L^{2/3}
    \right)^{3/2}, \quad \bar{\Delta}_L = \Delta_0\, \mathrm{e}^{-\pi \gamma_L
    /4}\,,
\end{equation}
which in turn depend on the magnetic field $H$ via parameter
\begin{equation}\label{gamma_L}
    \gamma_L = \bar{\Gamma}_L / \bar{\Delta}_L.
\end{equation}

The effective depairing energy $\bar{\Gamma}_L$ takes into account
the phase gradient created by the $L-$quantum vortex and the
effect of the magnetic field $H$, and can be calculated by the
averaging of the inhomogeneous depairing parameter
$\Gamma_L(r,\,H)$~(\ref{depair-par}),~(\ref{super_velocity}):
\begin{eqnarray}\label{GammaL-aver}
    \bar{\Gamma}_L(H,\,\mathcal{D}_L) &=&
        \frac{\pi \hbar}{D S_Z} \int\limits_0^R r\, \left[ 1 - Z_{\mathcal{D}_L}(r) \right]^2 v_L^2(r)\, dr\,, \\
     S_Z &=& 2\pi \int\limits_0^R r\, \left[ 1 - Z_{\mathcal{D}_L}(r) \right] dr \,.\nonumber
\end{eqnarray}
The relation~(\ref{GammaL-aver}) accounts for the presence of the
core of the $L-$quantum vortex via the factor $1 -
Z_{\mathcal{D}_L}(r)$ in the superfluid velocity $\left[ 1 -
Z_{\mathcal{D}_L}(r) \right] v_L(r)$ and the effective area of the
disk $S_Z$. As a result, the effective depairing energy
$\bar{\Gamma}_L$~(\ref{GammaL-aver}) and parameter $\gamma_L$~(\ref{gamma_L}) depend on the magnetic field $H$ and the length
scale $\mathcal{D}_L$. The kernel~(\ref{M-function}) takes the
form
\begin{multline}
    M_L(r,\,E,\,\epsilon) \simeq Z_{\mathcal{D}_L}(r)^2 + \left[1 - Z_{\mathcal{D}_L}(r)\right]^2
    \bar{M}_L(E,\,\epsilon) + \\ + Z_{\mathcal{D}_L}(r)\left[1 - Z_{\mathcal{D}_L}(r)\right]
    \bar{K}_L(E,\,\epsilon) \,, \label{M-function-aver}
\end{multline}
where the functions
\begin{eqnarray}
&&\bar{M}_L(E,\,\epsilon) = \bar{N}_L(E) \bar{N}_L(E+\epsilon) - \bar{B}_L(E) \bar{B}_L(E+\epsilon)\,, \label{M-core-aver} \\
&&\bar{K}_L(E,\,\epsilon) = \bar{N}_L(E) + \bar{N}_L(E+\epsilon)\,
\label{K-core-aver}
\end{eqnarray}
do not depend on radius $r$.

In the Meissner state ($L = 0$, $Z_0 = 0$) the depairing energy
$\bar{\Gamma}_0$ can be easily calculated by the averaging of the
inhomogeneous depairing parameter $\Gamma_0(r,\,H)$~(\ref{depair-par}),~(\ref{super_velocity}) over the radial
direction
\begin{equation}\label{Gamma0-aver}
    \bar{\Gamma}_0(H) = \frac{2}{R^2} \int\limits_0^R dr\, r\, \Gamma_0(r, H) =
        \frac{\hbar D}{4 R^2} \left(\frac{H}{H_0}\right)^2\,.
\end{equation}
The kernel $M_0(r,\,E,\,\epsilon)$ of the second integral
in~(\ref{Q_eph_gen_result1}) does not depend on radial coordinate
$r$ ($ M_0(r,\,E,\,\epsilon) \simeq \bar{M}_0(E,\,\epsilon)$) and
can be found by replacing the functions $N_0(r,\,E)$ and
$B_0(r,\,E)$ with $\bar{N}_0(E)$ and $\bar{B}_0(E)$, respectively,
in the expression~(\ref{M-function}). The comparison of the above
averaged description with the full numerical solution for the
density of states can be seen in Fig.~\ref{Fig-DOS}(e) where the
homogeneous density of states $\bar{N}_0(E) =
\mathrm{Re}[\cos\bar{\theta}_0]$ determined by the solution of
Eq.~\eqref{Usadel-aver_eq} is shown.

In the vortex state ($L \ne 0$) the unknown scale $\mathcal{D}_L$
depends on the disk geometry via the spatial distributions of the order
parameter $\Delta_L(r)$ and the local density of states
$N_L(r,\,E)$ in the disk, and in general $\mathcal{D}_L$ is a
function of temperature $T$, magnetic field $H$ and vorticity $L$.
In order to account for the reduction in the average order
parameter $\bar{\Delta}_L$ and the minigap $\bar{E}_\mathrm{g}$
which occurs with increasing $H$ for a fixed vorticity $L$, the
cutoff radius $\mathcal{D}_L$ is assumed to obey the following
relation
\begin{equation}\label{CutOff}
    \mathcal{D}_L(H) = a_L \xi_L(H)\,,
\end{equation}
where the coherence length
\begin{equation}\label{CoherLength}
    \xi_L(H) = \sqrt{\hbar D  / 2 \bar{E}_\mathrm{g}}
\end{equation}
plays the role of the characteristic length scale of the Green's
functions outside the vortex core, and $a_L$ is a fixed fitting
parameter for the $L-$th orbital mode.

Finally, the expressions~\eqref{Eg-Delta}-\eqref{GammaL-aver},~\eqref{CutOff}-\eqref{CoherLength}
define the implicit relation between the scale $\mathcal{D}_L$ and
magnetic field $H$, which after the substitution into the
expression~(\ref{GammaL-aver}) for the homogeneous depairing
energy $\bar{\Gamma}_L$ gives us the average normal ($\sim
\cos\bar{\theta}_L$) and anomalous ($\sim \sin\bar{\theta}_L$)
Green's functions from the solution of the algebraic equation~(\ref{Usadel-aver_eq}).

The general expressions~(\ref{Q_eph_gen_result1}) for the
electron-phonon heat flow can be significantly simplified in low
temperature limit $T_{\rm ph} \ll T \ll \bar{E}_g/k_B$, where we neglect the
terms $e^{-\bar{E}_g / k_B T_{\rm ph}}$ with respect to $e^{-\bar{E}_g
/ k_B T}$, using approximation of the kernel of the integrals
$M(r,\,E,\,\epsilon)$~(\ref{M-function-aver})-(\ref{K-core-aver}).
We calculate the coordinate--resolved electron-phonon heat flux
$\mathcal{P}_L(r)$ for any profile $Z_{\mathcal{D}_L}(r)$
combining the procedure, described in
Refs.~\cite{Kopnin-NeqSc,Timofeev-prl09}, and the solution of
Usadel equation~(\ref{Usadel-aver_eq}) (see Appendix~\ref{Append}
for details).

As a result in the leading approximation in $\bar{E}_g/(k_B T)$
for $T_{\rm ph} = 0$ and $T \ll \bar{\Gamma}^{2/3} \bar{E}_g^{1/3}/k_B$
the electron-phonon heat flow $\dot{Q}_L(r)$ within the central
part of the disk of the a radius $r \le R$ into the phonon bath
for the orbital mode $L$ is given by the expressions
\begin{equation}\label{Q_eph_gen_result1-Appr}
    \dot{Q}_L(r) =\frac{ \dot{Q}_N}{\pi R^2} \int\limits_0^r d r'\, r'\, \bar{\mathcal{P}}_L(r')\,,
\end{equation}
\begin{widetext}
\begin{multline}
   \bar{\mathcal{P}}_L(r) \approx 2\pi \left\{ Z_{\mathcal{D}_L}^2(r) +
    [1 - Z_{\mathcal{D}_L}(r)]^2 \frac{128}{189 \zeta(5)}\, \frac{k_B T}{\Gamma^{2/3} \bar{E}_g^{1/3}}
    \left[ 1 + \frac{21 \pi}{256} \left( \frac{\bar{E}_g}{k_B T} \right)^3 \mathrm{e}^{-\bar{E}_g / k_B T} \right]
                                \mathrm{e}^{-\bar{E}_g / k_B T} + \right. \\
        \left. + Z_{\mathcal{D}_L}(r) [1 - Z_{\mathcal{D}_L}(r)] \frac{\sqrt{\pi / 6}}{48 \zeta(5)}
            \left( \frac{\bar{E}_g \bar{\Delta}}{\bar{\Gamma}^2} \right)^{1/3}
            \left( \frac{\bar{E}_g}{k_B T} \right)^{7/2} \mathrm{e}^{-\bar{E}_g /  k_B T} \right\}\,,   \label{Q_eph_gen_result2-Appr}\
\end{multline}
\end{widetext}
where the first term in curvy brackets
in~\eqref{Q_eph_gen_result2-Appr} corresponds  to the normal core
contribution, the second term~---~to the superconducting part with
the hard gap, while the last term provides the cross-contribution
with the kernel~\eqref{K-core-aver}. $Q_N = \mathcal{V} \Sigma
T^5$ is the electron--phonon heat flow in the normal state of the
disk of volume $\mathcal{V} = \pi R^2 d$.

The results of the fitting for the trial function
$Z_{\mathcal{D}_L}(r) = \mathrm{e}^{-r^2 / \mathcal{D}_L^2}$ are
shown in Fig.~\ref{Fig4-Q(H)}. The best fits of
$\dot{Q}_L(R)$~(\ref{Q_eph_gen_result1-Appr}),~(\ref{Q_eph_gen_result2-Appr})
to the numerical simulations are obtained with $a_L \simeq
1.5;\,2.5;\,4.1;\,8.0$ for $L = 1 \div 4$, respectively. The inset
in Fig.~\ref{Fig4-Q(H)} shows the dependence of the value
$\bar{\Gamma}^{2/3} \bar{E}_g^{1/3}$ (solid lines) and the hard
gap $\bar{E}_g$ (dashed lines) determined by the
relations~(\ref{Eg-Delta}) on the magnetic flux $\phi$ for the
best fits. The condition $\bar{\Gamma}^{2/3} \bar{E}_g^{1/3} > k_B
T$ determines the values of the applied magnetic field $H \gtrsim
H_0$ for which the approximate
expressions~(\ref{Q_eph_gen_result1-Appr}),~(\ref{Q_eph_gen_result2-Appr})
are correct. One can see from Fig.~\ref{Fig4-Q(H)} that within the
validity range $\bar{\Gamma}^{2/3} \bar{E}_g^{1/3} > k_B T$ the
approximation agrees reasonably well with the exact result. In the
range $H<H_0$ of the Meissner regime, where the above validity
condition is violated, our semiquantitative model obviously fails
as it does not take into account highly inhomogeneous distribution
of Meissner currents (see, e.g., $P_0$ in Fig.~\ref{Fig-PQ}(a)).

Figures \ref{Fig-DOS}(e-h) provide the comparison of the exact
spatially resolved LDOS $N_L(r\,,E)$ vs energy $E$ (\ref{N_L-B_L})
shown in Figs.~\ref{Fig-DOS}(a-d) to the approximate
ones~(\ref{N_L-appr}) corresponding to the best fits for values of
the magnetic field $H = H_{s1}\,,H_{s2}$.

The corresponding spatially resolved heat flow density $\bar{\mathcal{P}}_L(r)$~(\ref{Q_eph_gen_result2-Appr})
and the heat flow $\dot{Q}_L(r)$~(\ref{Q_eph_gen_result1-Appr}) are shown in Fig.~\ref{Fig-PQ} by
the dashed lines.
One should notice the semiquantitative agreement between exact numerically calculated and
approximate curves, except for the cases Fig.~\ref{Fig-PQ}(a,~c), when the Meissner screening currents
essentially suppress the SC order parameter $\Delta(r)$ near the
disk edge, because the simple approximate model~(\ref{N_L-appr}),~(\ref{B_L-appr}) fails to account for this
suppression.

\begin{figure}[tb]
\includegraphics[width=0.45\textwidth]{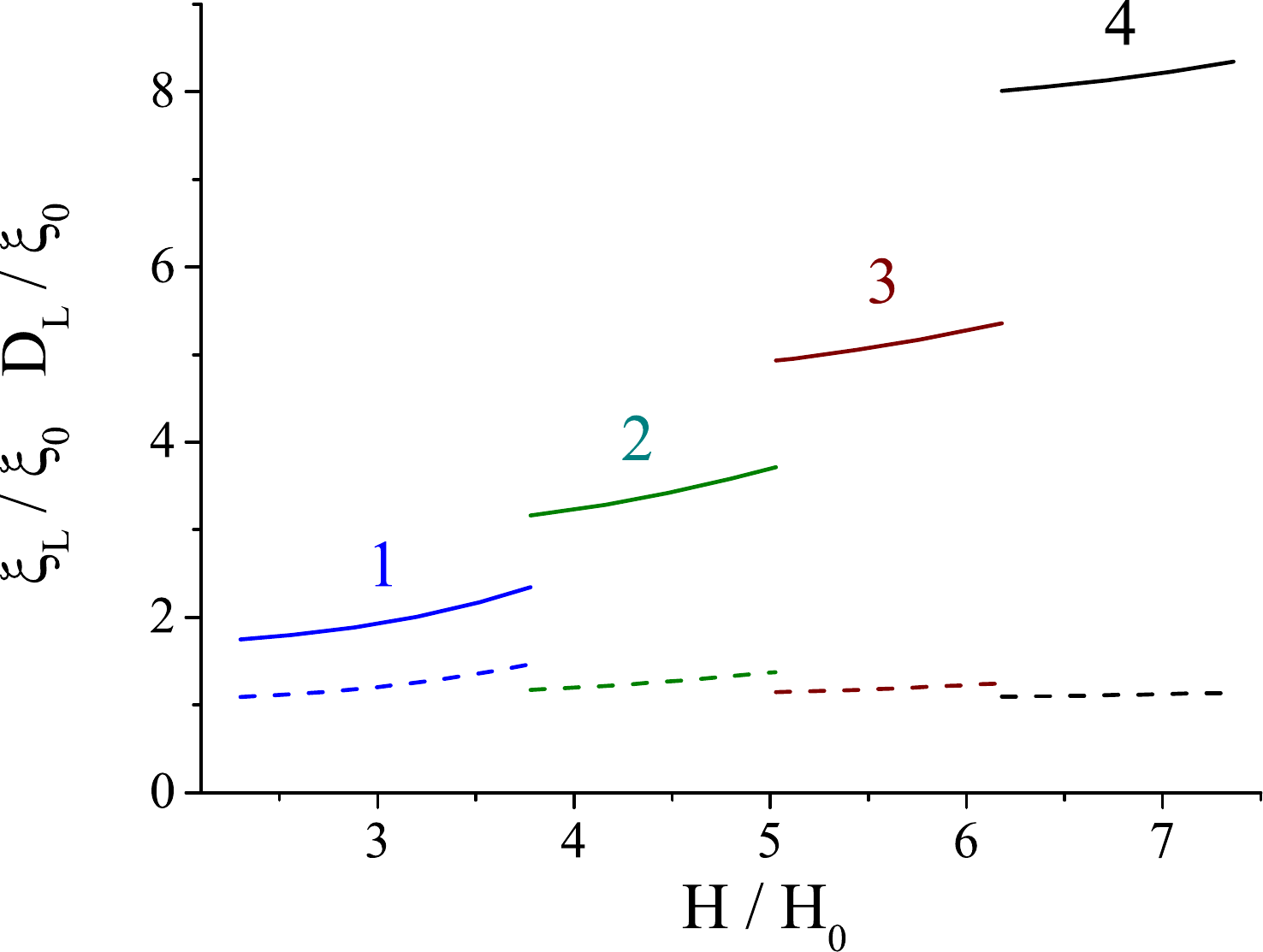}
\caption{(Color online) \textbf{The dependence of the scale
$\mathcal{D}_L$ (solid lines) and the coherence length $\xi_L$
(dashed lines) on the magnetic field $H$ in the SC disk} of the
radius $R = 4 \xi_0$ for the temperature $T = 0.1 T_{\rm cs}$. The
numbers near the curves denote the corresponding values of
vorticity $L$.} \label{Fig8-R(H)}
\end{figure}

Figure~\ref{Fig8-R(H)} shows the dependencies of
the coherence length $\xi_L(H)$ and the corresponding length scale $\mathcal{D}_L(H)$
on the magnetic field $H$ for different vorticities $L$. For a
fixed vorticity $L$ the coherence length $\xi_L(H)$~(\ref{CoherLength}) grows while sweeping the magnetic field up,
$H_{sL} \le H \le H_{s(L+1)}$, due to increase of the screening
currents in a small disk and shrinking of the gap
$\bar{E}_\mathrm{g}$. As a result, the radius $\mathcal{D}_{L}$
increases slightly within the branch $L$. The downward jumps in
the coherence length $\xi_L(H)$~(\ref{CoherLength}) at $H =
H_{sL}$ are caused by the increase in the hard spectral gap value
$\bar{E}_\mathrm{g}$ as the vortex enters the sample, see Fig.~\ref{Fig-DOS}. %
%

\section{Conclusions}\label{SumUpSection}

In conclusion, on the basis of the Usadel theory we have
calculated the electron--phonon heat transfer $\dot{Q}$ in a
diffusive mesoscopic SC disk of the size comparable to several
superconducting coherence lengths $\xi_0$,  placed in the external
magnetic field $H$ oriented perpendicular to the plane of the
disk. The strong confinement effects of the screening
supercurrents are responsible for the formation of the giant
($L-$quantum) vortices in the disk center. The giant vortex core
and the regions near the sample edge with the reduced spectral gap
form potential wells (traps) for quasiparticles responsible for
the heat transfer. We have demonstrated that the transitions
between the superconducting states with different vorticities $L$
provoke abrupt changes (jumps) in $\dot{Q}(H)$ attributed to the
entry/exit of vortices while sweeping the magnetic field. The
smooth growth of the electron--phonon heat flow  $\dot{Q}_L(H)$
takes place for a fixed vorticity $L$ while sweeping the magnetic
field up, due to the increase of the screening currents and
shrinking of the hard or soft spectral gap in the density of
states. We have shown that the electron--phonon heat flow in the
Meissner and vortex states of mesoscopic samples can be
effectively controlled by the  external magnetic field.
We develop the semiquantitative approximate model the description of the
quasiparticle trapping which works reasonably well beyond the Meissner state.
Our numerical analysis of the electron--phonon relaxation confirms
the validity and efficiency of the above
semiquantitative model. We expect our results can stimulate
further experimental work on the controllable QPs trapping in the
vortex state of mesoscopic superconductors and superconducting hybrid devices.

\acknowledgments 
We acknowledge Prof. Jukka P. Pekola for fruitful discussions.
This work was supported,  by the
Center of Excellence ``Center of Photonics''
funded by The Ministry of Science and Higher Education of the Russian
Federation, contract 075-15-2020-906.

\onecolumngrid
\appendix

\section{Homogeneous approximation in a mesoscopic disk}
\label{Append}
In this section we derive the expression for the
electron-phonon heat flow $\dot Q_{eph}$~(\ref{Q_eph_gen_result1})-(\ref{N_L-B_L}) in a mesoscopic
superconducting disk with a $L$-quantum vortex in the center within the approximation~\eqref{N_L-appr},~\eqref{B_L-appr}.
Using the kernel~(\ref{M-function-aver}), we decompose the heat flow density~(\ref{Q_eph_gen_result1}) into three parts as follows
\begin{equation}\label{PL_homo}
    \mathcal{P}_L(r) = \Pi^{(1)} F^2(r) + \Pi^{(2)} [1 - F(r)]^2 + \Pi^{(3)} F(r)\,[1 - F(r)] \,,
\end{equation}
where the coefficients $\Pi^{(i)}$ are given by the unified expression
%
\begin{equation}
   \Pi_L^{(i)} = \frac{\Sigma}{24 \zeta(5) k_B^5} \int\limits_{0}^{\infty}\epsilon^3 \left[\, n_T(\epsilon)
            - n_{T_{\rm ph}}(\epsilon)\, \right] d\epsilon \int\limits_{-\infty}^{\infty}
             \mathcal{M}_L^{(i)}(E,\,\epsilon)\,\left[\, f_T(E) - f_T(E+\epsilon)\,\right] dE \,, \label{Pi_homo}
\end{equation}
%
with the kernels $\mathcal{M}_L^{(1)}(E,\,\epsilon) =  1$,
$\mathcal{M}_L^{(2)}(E,\,\epsilon) = \bar{M}_L(E,\,\epsilon)$, and
$\mathcal{M}_L^{(3)}(E,\,\epsilon) = \bar{K}_L(E,\,\epsilon)$, respectively.
The coefficient
\begin{equation}
   \Pi^{(1)} = \frac{\Sigma}{24 \zeta(5) k_B^5} \int\limits_{0}^{\infty}\epsilon^3 \left[\, n_T(\epsilon)
            - n_{T_{\rm ph}}(\epsilon)\, \right] d\epsilon \int\limits_{-\infty}^{\infty}
             \left[\, f_T(E) - f_T(E+\epsilon)\,\right] dE = \Sigma\, [ T^5 - T_{\rm ph}^5] \,, \label{Pi1_homo}
\end{equation}
describes the heat flow density in the normal state and determines
the contribution of the vortex core.

For simplicity the further calculations are done in low temperature limit $T_0 \ll T\ll
\bar{E}_g/k_B$ in order to neglect the temperature dependence of the gap
and the order parameter and the terms $e^{-\bar{E}_g/k_B T_{\rm ph}}$
with respect to $e^{-\bar{E}_g/k_B T}$.
In this case the values
$\Pi^{(i)}$~(\ref{Pi_homo}) do not depend on the phonon temperature
$T_{\rm ph}$ and  $\Pi^{(1)} \approx \Sigma\, T^5$.

In the low
temperature limit the main contributions to the coefficients
$\Pi^{(2)}$ and $\Pi^{(3)}$ arise from the vicinity $0 < |E| -
\bar{E}_g \lesssim k_B T$ of the hard gap value $\pm
\bar{E}_g$, leading to the smallness of the positive parameter $\delta E
= |E| - \bar{E}_g \ll \bar{E}_g$.
The expansion of $\bar{N}_L(E)$ and $\bar{B}_L(E)$ over $\delta E$
takes the form
\begin{equation}
    \bar{N}_L^2(E) \simeq \Theta(\delta E) \frac{2 \delta E\, \bar{\Delta}_L^{2/3}}{3 \bar{\Gamma}_L^{4/3}\,\bar{E}_g^{1/3}}\,, \qquad
    \frac{\bar{B}_L(E)}{\bar{N}_L(E)\, \mathrm{sign}(E)} \simeq \left( \frac{\bar{E}_g}{\bar{\Delta}_L}\right)^{1/3}\,. \label{NLBL-expans}
\end{equation}
Here $\Theta(x)$ is the Heaviside theta-function. Substituting~(\ref{NLBL-expans})
into expressions~(\ref{M-core-aver}, \ref{K-core-aver}) we obtain the final expression
for the kernels $\bar{M}_L(E,\,\epsilon)$ and $\bar{K}_L(E,\,\epsilon)$
\begin{equation}
    \bar{M}_L(E,\,\epsilon) \simeq \frac{2 \sqrt{\delta E\, \delta E^\prime}\,\Theta(\delta E)\,\Theta(\delta E^\prime)}{3
    \bar{\Gamma}_L^{2/3}\,\bar{E}_g^{1/3}}\,, \qquad
    \bar{K}_L(E,\,\epsilon) \simeq \sqrt{\frac{2 \bar{\Delta}_L^{2/3}}{3 \bar{\Gamma}_L^{4/3}\,\bar{E}_g^{1/3}}}
    \left[ \sqrt{\delta E}\,\Theta(\delta E) + \sqrt{\delta E^\prime}\,\Theta(\delta E^\prime) \right]\,,
\end{equation}
which are correct for $T \ll \bar{\Gamma}^{2/3} \bar{E}_g^{1/3}/k_B$,
where $\delta E^\prime = |E + \epsilon| - \bar{E}_g$. Splitting
the integration in $\Pi^{(2)}$~(\ref{Pi_homo}) into positive and
negative parts in the electronic energy $E$ and taking into
account that $\bar{M(}E,\,\epsilon) = 0$ if any of two conditions
$|E| < \bar{E}_g$ and $|E + \epsilon| < \bar{E}_g$ is valid, the
coefficient $\Pi^{(2)}$ can be simplified in the low temperature
limit as follows
\begin{equation}
   \Pi^{(2)} \approx \frac{\Sigma}{24 \zeta(5) k_B^5} \left[ 2 \int\limits_{0}^{\infty}\epsilon^3
        \mathrm{e}^{-\epsilon/k_B T} d\epsilon \int\limits_{\bar{E}_g}^{\infty}
             \bar{M}_L(E,\,\epsilon)\,\mathrm{e}^{-E / k_B T} dE + \int\limits_{2 \bar{E}_g}^{\infty}\epsilon^3 \mathrm{e}^{-\epsilon/k_B T} d\epsilon \int\limits_{\bar{E}_g - \epsilon}^{-\bar{E}_g} \bar{M}_L(E,\,\epsilon)\,dE \right]\,, \label{Pi2_homo}
\end{equation}
As a result taking into account only the leading terms in the small parameter $k_B T / \bar{E}_g$
we obtain
\begin{gather}\label{Pi2_res}
    \Pi^{(2)} = \frac{128\, \Sigma T^5}{189\, \zeta(5)}\,
        \frac{k_B T}{\bar{\Gamma}_L^{2/3} \bar{E}_q^{1/3}} \left[ 1 +
        \frac{21 \pi}{256} \left( \frac{\bar{E}_g}{k_B T} \right)^3
        \mathrm{e}^{-\bar{E}_g / k_B T} \right] \mathrm{e}^{-\bar{E}_g / k_B T} \,.
\end{gather}
In order to simplify the expression for $\Pi^{(3)}$ with $E \ge \bar{E}_g$, we neglect $n_{T_{\rm ph}}(\epsilon)$ and use the following equalities
$$
    n_T(\epsilon)\, f_T(E-\epsilon) = f_T(E)\left[ n_T(\epsilon) + f_T(E-\epsilon)\right]\,, \qquad
    n_T(\epsilon)\, f_T(E+\epsilon) = f_T(E)\left[ n_T(\epsilon) - f_T(E+\epsilon)\,
        \mathrm{e}^{E/k_B T} \right] \,,
$$
As a result we obtain
\begin{equation}
   \Pi^{(3)} \approx \frac{\Sigma}{12 \zeta(5) k_B^5} \int\limits_{0}^{\infty}\epsilon^3
        n_T(\epsilon) d\epsilon \int\limits_{\bar{E}_g}^{\infty}
             \bar{N}_L(E)\,\left[ f_T(E-\epsilon) + f_T(E+\epsilon)\,
        \mathrm{e}^{E/k_B T} \right]\,, \label{Pi2_homo}
\end{equation}
leading to
\begin{equation}
   \Pi^{(3)} \approx \frac{\sqrt{\pi/6}\, \Sigma T^5}{48 \zeta(5)}
     \left( \frac{\bar{E}_g \bar{\Delta}_L}{\bar{\Gamma}_L^2}\right)^{1/3}
    \left( \frac{\bar{E}_g}{k_B \bar{T}} \right)^{7/2} \mathrm{e}^{-\bar{E}_g / k_B \bar{T}} \,. \label{Pi3_homo}
\end{equation}

Finally, substituting~(\ref{PL_homo}),~(\ref{Pi1_homo}),~(\ref{Pi2_homo}),~(\ref{Pi3_homo})
into the expressions~(\ref{Q_eph_gen_result1}), one obtains
simplified expressions~(\ref{Q_eph_gen_result1-Appr}),~(\ref{Q_eph_gen_result2-Appr}) for
the electron-phonon heat flow $\dot{Q}_L(r)$ into the phonon bath
for the orbital mode $L$.

\twocolumngrid

\end{document}